\documentclass[reqno,12pt]{article}
\usepackage{ae} 
\usepackage[T1]{fontenc}
\usepackage[ansinew]{inputenc}
\usepackage{amsmath}
\usepackage{amssymb}
\usepackage{graphicx}
\usepackage{bm}
\usepackage{caption}
\usepackage{subcaption}
\usepackage{dsfont}
\usepackage{mdframed}
\usepackage{mathtools}
\usepackage{lipsum}
\usepackage{color}
\definecolor{darkblue}{cmyk}{0.9,0.9,0,0}
\usepackage[colorlinks=true,linkcolor=darkblue,citecolor=darkblue,urlcolor=darkblue]{hyperref}
\usepackage{epsfig}

\usepackage[dvipsnames]{xcolor}
\usepackage{graphicx}

\newmuskip\pFqmuskip

\newcommand*\pFq[6][8]{%
  \begingroup 
  \pFqmuskip=#1mu\relax
  \mathcode`\,=\string"8000
  \begingroup\lccode`\~=`\,
  \lowercase{\endgroup\let~}\pFqcomma
  {}_{#2}F_{#3}{\left[\genfrac..{0pt}{}{#4}{#5};#6\right]}%
  \endgroup
}
\newcommand{\pFqcomma}{\mskip\pFqmuskip}

\newcommand{\beq}{\begin{equation}}
\newcommand{\eeq}{\end{equation}}
\newcommand\beqa{\begin{eqnarray}}
\newcommand\eeqa{\end{eqnarray}}
\newcommand\bea{\begin{array}}
\newcommand\eea{\end{array}}

\def\XXint#1#2#3{{\setbox0=\hbox{$#1{#2#3}{\int}$}
\vcenter{\hbox{$#2#3$}}\kern-.5\wd0}}

\newcommand{\neqa}{\nonumber\end{eqnarray}}
\newcommand{\la}[1]{\label{#1}}

\renewcommand{\d}{\partial}

\newcommand{\<}{{\langle}}
\renewcommand{\>}{{\rangle}}

\newcommand{\re}{\relax{\rm I\kern-.18em R}}

\renewcommand{\sp}{p\hspace{-.40em}/}

\def\su2{{SU(2)}}

\def\[{\left[}
\def\]{\right]}

\def\({\left(}
\def\){\right)}
\def\[{\left[}
\def\]{\right]}

\def\<{\langle}
\def\>{\rangle}

\def\i2{\frac{i}{2}}

\def\spi{\relax{\rm \pi\kern-0.5em /}}
\def\sA{\relax{\rm A\kern-0.5em /}}
\def\sp{\relax{\rm p\kern-0.5em /}}
\def\sd{\relax{\rm \d\kern-0.5em /}}
\def\sk{\relax{\rm k\kern-0.5em /}}
\def\sn{\relax{\rm n\kern-0.5em /}}
\def\sl{\relax{\rm l\kern-0.5em /}}
\def\sP{\relax{\rm P\kern-0.7em /}}
\def\sBethe{\relax{\rm \Bethe\kern-0.5em /}}

\usepackage{varioref}
\usepackage{makeidx}
\makeindex

\newcommand\blfootnote[1]{%
  \begingroup
  \renewcommand\thefootnote{}\footnote{\hspace{-6mm}#1}%
  \addtocounter{footnote}{-1}%
  \endgroup
}

\usepackage[english]{babel}
\begin{document}


\thispagestyle{empty}

\renewcommand{\thefootnote}{\fnsymbol{footnote}}
\setcounter{page}{1}
\setcounter{footnote}{0}
\setcounter{figure}{0}

\vspace{-0.4in}

\begin{center}
$$$$
{\Large\textbf{\mathversion{bold}
Spinning Hexagons}\par}
\vspace{1.0cm}

\textrm{Carlos Bercini$^\text{\tiny 1}$, Vasco Gon\c{c}alves$^\text{\tiny 1,\tiny 2}$, Alexandre Homrich$^\text{\tiny 3}$, Pedro Vieira$^\text{\tiny 1,\tiny 3}$}
\blfootnote{\tt  \#@gmail.com\&/@\{carlos.bercini,vasco.dfg,alexandre.homrich,pedrogvieira\}}
\\ \vspace{1.2cm}
\footnotesize{\textit{
$^\text{\tiny 1}$ICTP South American Institute for Fundamental Research, IFT-UNESP, S\~ao Paulo, SP Brazil 01440-070 \\
$^\text{\tiny 2}$Centro de Fisica do Porto e Departamento de Fisica e Astronomia, Faculdade de Ciencias da Universidade do Porto, Porto 4169-007, Portugal   \\
$^\text{\tiny 3}$Perimeter Institute for Theoretical Physics,
Waterloo, Ontario N2L 2Y5, Canada}
\vspace{4mm}
}
\end{center}

\par\vspace{1.5cm}


\vspace{2mm}
\begin{abstract}
We reduce the computation of three point function of three spinning operators with arbitrary polarizations to a statistical mechanics problem via the hexagon formalism. The central building block of these correlation functions is the \textit{hexagon partition function}. We explore its analytic structure and use it to generate perturbative data for spinning three point functions. For certain polarizations and any coupling, we express the full asymptotic three point function in determinant form. With the integrability approach established we open the ground to study the large spin limit where dualities with null Wilson loops and integrable pentagons must appear. 
\end{abstract}

\newpage

\setcounter{page}{1}
\renewcommand{\thefootnote}{\arabic{footnote}}
\setcounter{footnote}{0}



{
\tableofcontents
}



\newpage

\section{Introduction} 
Three point functions of single trace operators in planar $\mathcal{N}=4$ SYM describe the scattering of three closed strings in AdS and are thus given by a pairs of pants. Pairs of pants can be obtained by stitching two hexagons together. That is how tailors make pants and it is also how one computes three point functions in this gauge theory using integrability \cite{hexagons}. 
This paper is about spinning pair of pants where two or more operators have spin. In this case the 3pt function is given by a sum of conformal invariant tensor structures and we need to explain how the hexagons extract the coefficient multiplying each such structure.  For three twist-two operators with spin 2, 4 and 6, for instance, we have the perturbative result 
\cite{Multi}
\begingroup\makeatletter\def\f@size{7}\check@mathfonts
\def\maketag@@@#1{\hbox{\m@th\large\normalfont#1}}%
\begin{align}
C(2,4,6) &= \frac{1}{84\sqrt{55}}\Bigg(\langle 1  
     1\rangle  \langle 1  
     3\rangle  \langle 2  
     2\rangle^3  \langle 2  
     3\rangle  \langle 3  
     1\rangle  \langle 3  
     2\rangle  \langle 3  
     3\rangle^4 \left(\textcolor{blue}{960} - \textcolor{Bittersweet}{\frac{2400756}{385}}g^2\right)+\langle 1  
     2\rangle^2  \langle 2  
     1\rangle^2  \langle 2  
     2\rangle^2  \langle 3  
     3\rangle^6 \left(\textcolor{blue}{15} - \textcolor{Bittersweet}{\frac{88797}{770}}g^2\right)+ \nonumber \\ 
     &+ \langle 1  
     1\rangle  \langle 1  
     3\rangle  \langle 2  
     3\rangle^4  \langle 3  
     1\rangle  \langle 3  
     2\rangle^4  \langle 3  
     3\rangle  \left(-\textcolor{blue}{2007} - \textcolor{Bittersweet}{\frac{822990467}{34650}}g^2\right)+\langle 1  
     3\rangle^2  \langle 2  
     2\rangle  \langle 2  
     3\rangle^3  \langle 3  
     1\rangle^2  \langle 3  
     2\rangle^3  \langle 3  
     3\rangle\left(-\textcolor{blue}{7680} - \textcolor{Bittersweet}{\frac{3113024}{231}}g^2\right)+ \nonumber \\
     &+ \langle 1  
     2\rangle  \langle 1  
     3\rangle  \langle 2  
     1\rangle  \langle 2  
     3\rangle^3  \langle 3  
     1\rangle  \langle 3  
     2\rangle^3  \langle 3  
     3\rangle^2 \left(-\textcolor{blue}{15360} - \textcolor{Bittersweet}{\frac{13877312}{1155}}g^2\right)+\langle 1  
     3\rangle^2  \langle 2  
     2\rangle^4  \langle 3  
     1\rangle^2  \langle 3  
     3\rangle^4\left(\textcolor{blue}{23} - \textcolor{Bittersweet}{\frac{7780337}{34650}}g^2\right) + \nonumber \\
     &+\langle 1  
     2\rangle  \langle 1  
     3\rangle  \langle 2  
     1\rangle  \langle 2  
     2\rangle  \langle 2  
     3\rangle^2  \langle 3  
     1\rangle  \langle 3  
     2\rangle^2  \langle 3  
     3\rangle^3 \left(\textcolor{blue}{23040} - \textcolor{Bittersweet}{\frac{12990112}{1155}}g^2\right) + \langle 1  
     1\rangle^2  \langle 2  
     2\rangle^4  \langle 3  
     3\rangle^6 \left(\textcolor{blue}{1} - \textcolor{Bittersweet}{\frac{814939}{34650}}g^2\right)+ \nonumber \\
     &+\langle 1  
     1\rangle  \langle 1  
     2\rangle  \langle 2  
     1\rangle  \langle 2  
     2\rangle^2  \langle 2  
     3\rangle  \langle 3  
     2\rangle  \langle 3  
     3\rangle^5 \left(\textcolor{blue}{576} - \textcolor{Bittersweet}{\frac{5341948}{1925}}g^2\right) + \langle 1  
     1\rangle^2  \langle 2  
     3\rangle^4  \langle 3  
     2\rangle^4  \langle 3  
     3\rangle^2 \left(\textcolor{blue}{327} + \textcolor{Bittersweet}{\frac{128149187}{34650}}g^2\right) + \nonumber \\
     &+\langle 1  
     3\rangle^2  \langle 2  
     2\rangle^3  \langle 2  
     3\rangle  \langle 3  
     1\rangle^2  \langle 3  
     2\rangle  \langle 3  
     3\rangle^3  \left(-\textcolor{blue}{1883} - \textcolor{Bittersweet}{\frac{14853709}{4950}}g^2\right)+\langle 1  
     1\rangle^2  \langle 2  
     2\rangle^2  \langle 2  
     3\rangle^2  \langle 3  
     2\rangle^2  \langle 3  
     3\rangle^4 \left(\textcolor{blue}{360} - \textcolor{Bittersweet}{\frac{3289708}{1155}}g^2\right)+ \nonumber \\
     &+\langle 1  
     2\rangle^2  \langle 2  
     1\rangle^2  \langle 2  
     2\rangle  \langle 2  
     3\rangle  \langle 3  
     2\rangle  \langle 3  
     3\rangle^5  \left(-\textcolor{blue}{567} - \textcolor{Bittersweet}{\frac{3484307}{1650}}g^2\right)+ \langle 1  
     1\rangle^2  \langle 2  
     2\rangle  \langle 2  
     3\rangle^3  \langle 3  
     2\rangle^3  \langle 3  
     3\rangle^3 \left(-\textcolor{blue}{640} - \textcolor{Bittersweet}{\frac{5365984}{3465}}g^2\right)+ \nonumber \\
     &+\langle 1  
     2\rangle  \langle 1  
     3\rangle  \langle 2  
     1\rangle  \langle 2  
     2\rangle^3  \langle 3  
     1\rangle  \langle 3  
     3\rangle^5 \left(\textcolor{blue}{192} + \textcolor{Bittersweet}{\frac{145072}{5775}}g^2\right)+\langle 1  
     1\rangle  \langle 1  
     2\rangle  \langle 2  
     1\rangle  \langle 2  
     2\rangle^3  \langle 3  
     3\rangle^6 \left(-\textcolor{blue}{16} + \textcolor{Bittersweet}{\frac{2405402}{17325}}g^2\right)+ \nonumber \\
     &+\langle 1  
     1\rangle  \langle 1  
     3\rangle  \langle 2  
     2\rangle^4  \langle 3  
     1\rangle  \langle 3  
     3\rangle^5  \left(-\textcolor{blue}{24} + \textcolor{Bittersweet}{\frac{1432546}{5775}}g^2\right)+ 
  \langle 1  
     1\rangle  \langle 1  
     3\rangle  \langle 2  
     2\rangle  \langle 2  
     3\rangle^3  \langle 3  
     1\rangle  \langle 3  
     2\rangle^3  \langle 3  
     3\rangle^2  \left(\textcolor{blue}{7680} + \textcolor{Bittersweet}{\frac{20817136}{1155}}g^2\right) + \nonumber \\
     &+\langle 1  
     3\rangle^2  \langle 2  
     3\rangle^4  \langle 3  
     1\rangle^2  \langle 3  
     2\rangle^4  \left(\textcolor{blue}{960} + \textcolor{Bittersweet}{\frac{272704}{231}}g^2\right) + 
  \langle 1  
     1\rangle  \langle 1  
     2\rangle  \langle 2  
     1\rangle  \langle 2  
     2\rangle  \langle 2  
     3\rangle^2  \langle 3  
     2\rangle^2  \langle 3  
     3\rangle^4 \left(-\textcolor{blue}{2880} + \textcolor{Bittersweet}{\frac{3467144}{1155}}g^2\right) + \nonumber \\
     &+\langle 1  
     2\rangle  \langle 1  
     3\rangle  \langle 2  
     1\rangle  \langle 2  
     2\rangle^2  \langle 2  
     3\rangle  \langle 3  
     1\rangle  \langle 3  
     2\rangle  \langle 3  
     3\rangle^4  \left(-\textcolor{blue}{5760} + \textcolor{Bittersweet}{\frac{1434656}{385}}g^2\right) + 
  \langle 1  
     2\rangle^2  \langle 2  
     1\rangle^2  \langle 2  
     3\rangle^2  \langle 3  
     2\rangle^2  \langle 3  
     3\rangle^4 \left(\textcolor{blue}{1440} + \textcolor{Bittersweet}{\frac{6762608}{1155}}g^2\right) + \nonumber \\
     &+ \langle 1  
     3\rangle^2  \langle 2  
     2\rangle^2  \langle 2  
     3\rangle^2  \langle 3  
     1\rangle^2  \langle 3  
     2\rangle^2  \langle 3  
     3\rangle^2  \left(\textcolor{blue}{8640} + \textcolor{Bittersweet}{\frac{852608}{77}}g^2\right) + 
  \langle 1  
     1\rangle  \langle 1  
     2\rangle  \langle 2  
     1\rangle  \langle 2  
     3\rangle^3  \langle 3  
     2\rangle^3  \langle 3  
     3\rangle^3  \left(\textcolor{blue}{2560} + \textcolor{Bittersweet}{\frac{46393456}{3465}}g^2\right)+ \nonumber \\
     &+\langle 1  
     1\rangle  \langle 1  
     3\rangle  \langle 2  
     2\rangle^2  \langle 2  
     3\rangle^2  \langle 3  
     1\rangle  \langle 3  
     2\rangle^2  \langle 3  
     3\rangle^3  \left(-\textcolor{blue}{5760} + \textcolor{Bittersweet}{\frac{19842568}{1155}}g^2\right)+ 
  \langle 1  
     1\rangle^2  \langle 2  
     2\rangle^3  \langle 2  
     3\rangle  \langle 3  
     2\rangle  \langle 3  
     3\rangle^5  \left(-\textcolor{blue}{48} + \textcolor{Bittersweet}{\frac{1389824}{1925}}g^2\right)+ \nonumber \\
     &+\left(\langle 1  
     1\rangle ^2 \langle 2  
     3\rangle ^4 \langle 3  
     2\rangle ^4 \langle 3  
     3\rangle ^2-\langle 1  
      1\rangle  \langle 1  
     3\rangle  \langle 2  
     3\rangle ^4 \langle 3  
     1\rangle  \langle 3  
     2\rangle ^4 \langle 3  
     3\rangle\right) \left(-\textbf{\textcolor{blue}{87}} + \textcolor{cyan}{\frac{\textbf{121910653}}{\textbf{34650}}}g^2\right)+ \nonumber \\
     & +\left( \langle 1  
     3\rangle ^2 \langle 2  
     2\rangle ^4 \langle 3  
     1\rangle ^2 \langle 3  
     3\rangle ^4-\langle 1  3\rangle ^2 \langle 2 
      2\rangle ^3 \langle 2  
     3\rangle  \langle 3  
     1\rangle ^2 \langle 3  
     2\rangle  \langle 3  
     3\rangle ^3\right)  \left(\textcolor{blue}{\textbf{37}} + \textcolor{cyan}{\frac{\textbf{15349637}}{\textbf{34650}}}g^2\right)+ \nonumber \\
     & + \left( \langle 1  
     2\rangle ^2 \langle 2  
     1\rangle ^2 \langle 2  
     2\rangle ^2 \langle 3  
     3\rangle ^6-\langle 1  2\rangle ^2 \langle 2 
      1\rangle ^2 \langle 2  
     2\rangle  \langle 2  
     3\rangle  \langle 3  
     2\rangle  \langle 3  
     3\rangle ^5\right) \left(\textcolor{blue}{\textbf{9}} + \textcolor{cyan}{\frac{\textbf{3091483}}{\textbf{11550}}}g^2\right)\Bigg)+\dots
   \label{C246}
\end{align}\endgroup
where $\langle ij \rangle
$ is a scalar product involving a left spinor parametrizing operator $i$ and a right spinor parametrizing operator $j$.\footnote{These spinor variables render the various conformal invariant tensor structures into simple scalar products, recalled in appendix \ref{spinors}.} A main goal of this paper is to develop the formalism to reproduce such results from integrability.

Not all terms are equally easy to get. The terms in blue, for instance, are the tree level contributions; we will develop an efficient recursion algorithm which will allow us to determine them all (and produce a plethora of new predictions for structure constants of larger twist operators). The boldfaced terms are what we call the \textit{abelian} terms; these structure constants are very integrability friendly as they lack a complicated so-called \textit{hexagon matrix part}; these abelian terms we can actually compute easily at one loop (in cyan here) or even at higher loops. The remaining non boldfaced orange terms are non-abelian one loop terms; we can also get them but it is quite painful to do so specially for operators of larger spin. 

At the center of all these integrability based computations is a beautiful partition function represented in figure \ref{triangle}. We call it the \textit{hexagon partition function}. The vertex in this partition function is Beisert's centrally extended $SU(2|2)$ R-matrix \cite{Beisert1} while the open boundary conditions are given by contracting each boundary with a fixed two dimensional spinor. The various possible choices of such spinors parametrize the various tensor structures described above. (For some choice of boundary conditions this partition function trivializes -- leading to the much simpler abelian contributions described above.) This hexagon partition function is the building block of all spinning correlators. It is an interesting object on its own right which we want to advertise in this paper. It has a beautiful very rich integrable structure, the surface of which we are barely starting to scratch. 

This paper is naturally split into two main sections. 

In section \ref{TriangleSec} we study the hexagon partition function mentioned above. This section might be interesting for hardcore integrabilists, even those with no interest in three point functions in SYM. We will study this partition function at weak coupling when it reduces to a rational vertex model partition function and at finite coupling where it is richer, of Hubbard type. Recursion relations, analytic continuations and several other tricks will play a key role in this analysis.

In section \ref{SecCtoH} we describe how to introduce polarizations into the hexagon formalism to compute spinning structure constants. We will do it mostly in the so called asymptotic regime where mirror particles can be ignored except in section \ref{CtoH2loops} where we perform some checks involving mirror particles at two loops. We will show how the triangle partition function naturally shows up and the fundamental building block and we will test the hexagon construction against perturbative data for a few simple examples with low spins.
  
\section{The Hexagon Partition Function}  \la{TriangleSec}
\subsection{The Partition Function}

\begin{figure}[t!]
\centering
\includegraphics[width=1\textwidth]{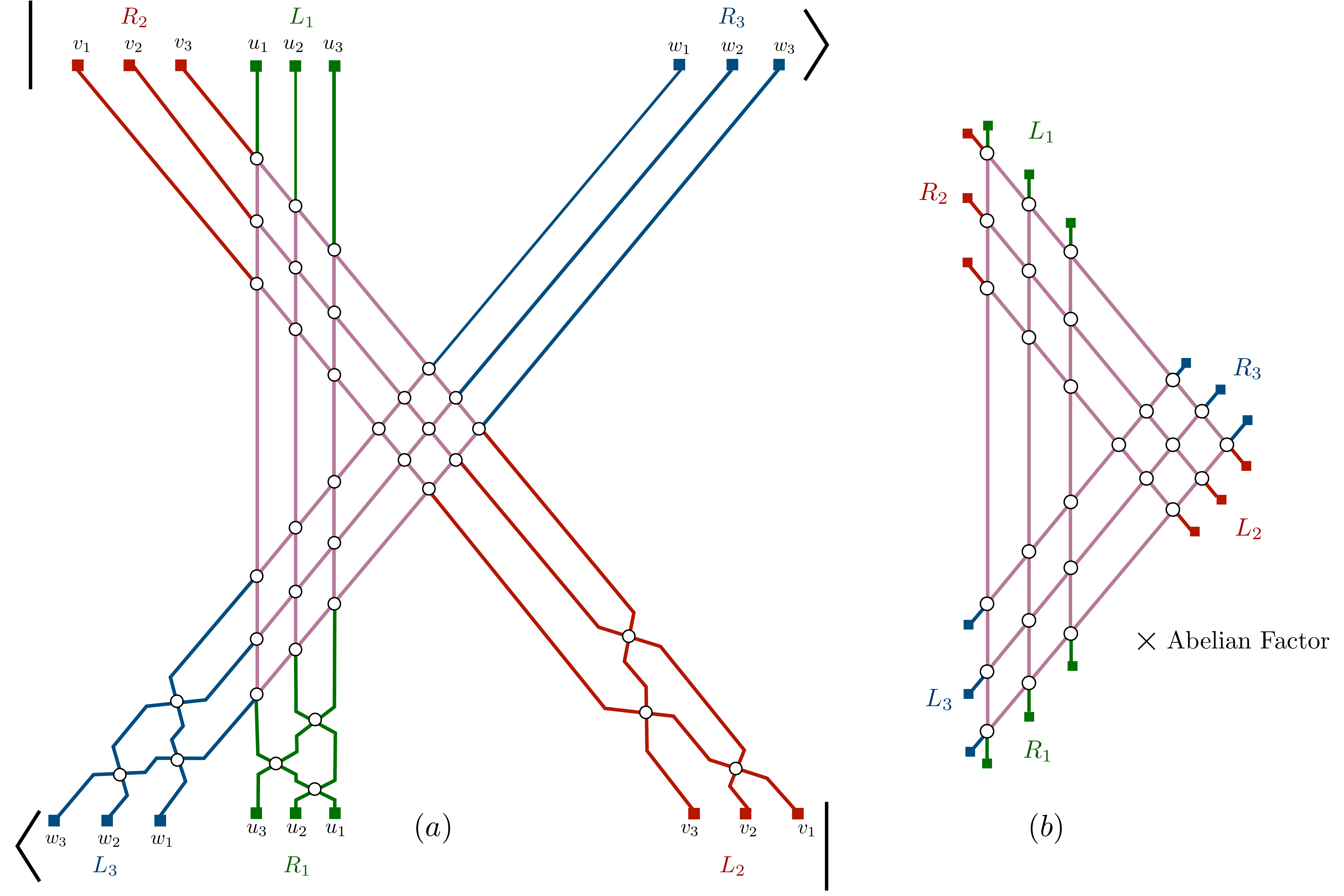}
\caption{(a) The hexagon partition function $Z_{J_1, J_2, J_3}$, illustrated here in the case $Z_{3,3,3}$, describe the scattering of three sets of fermions, labelled by their rapidities  $v_i, u_j, w_k$, in the $\mathcal{N} =4$ SYM spin-chain. Each set starts polarized in a fixed direction labeled by the spinors $R_2, L_1,R_3$ respectively. The particles then scatter in all possible pairings according to Beisert's $PSU(2|2)$ vertex. The final state is then projected into fermions of definite polarization spinors $L_2, R_1, L_3$.  In the gauge theory, the boundary conditions are set by the spacetime polarizations of spinning operators whose structure constants are governed by the hexagon partition function. Because the vertex is proportional to the identity when the incoming or outgoing particles are fermions with identical polarizations, the outer parts of the graph are trivial and result in a simple abelian factor. We represent these trivial scattering by the green red and blue colours. They can be factored out resulting in the equivalent representation (b). The interesting dynamics happens in the pink region in which particles from different sets interact.  }
\label{triangle}
\end{figure}

The central object in this paper is dubbed the \textbf{hexagon partition function} or simply the \textbf{hexagon}. It is defined in figure \ref{triangle}a. The name hexagon becomes clear when we realize it is given by a simpler partition function depicted in figure \ref{triangle}b.

As usual, the partition function can be thought of as 2d classical statistical mechanical model where we sum over statistical weights at each vertex (more precisely the vertex is the Shastry R-matrix of the Hubbard model) and where at the edges we impose appropriate boundary conditions. For the hexagon partition function there are six different boundary conditions to impose each parametrized by its own two dimensional spinor.

\begin{figure}[t!]
\centering
\includegraphics[width=1\textwidth]{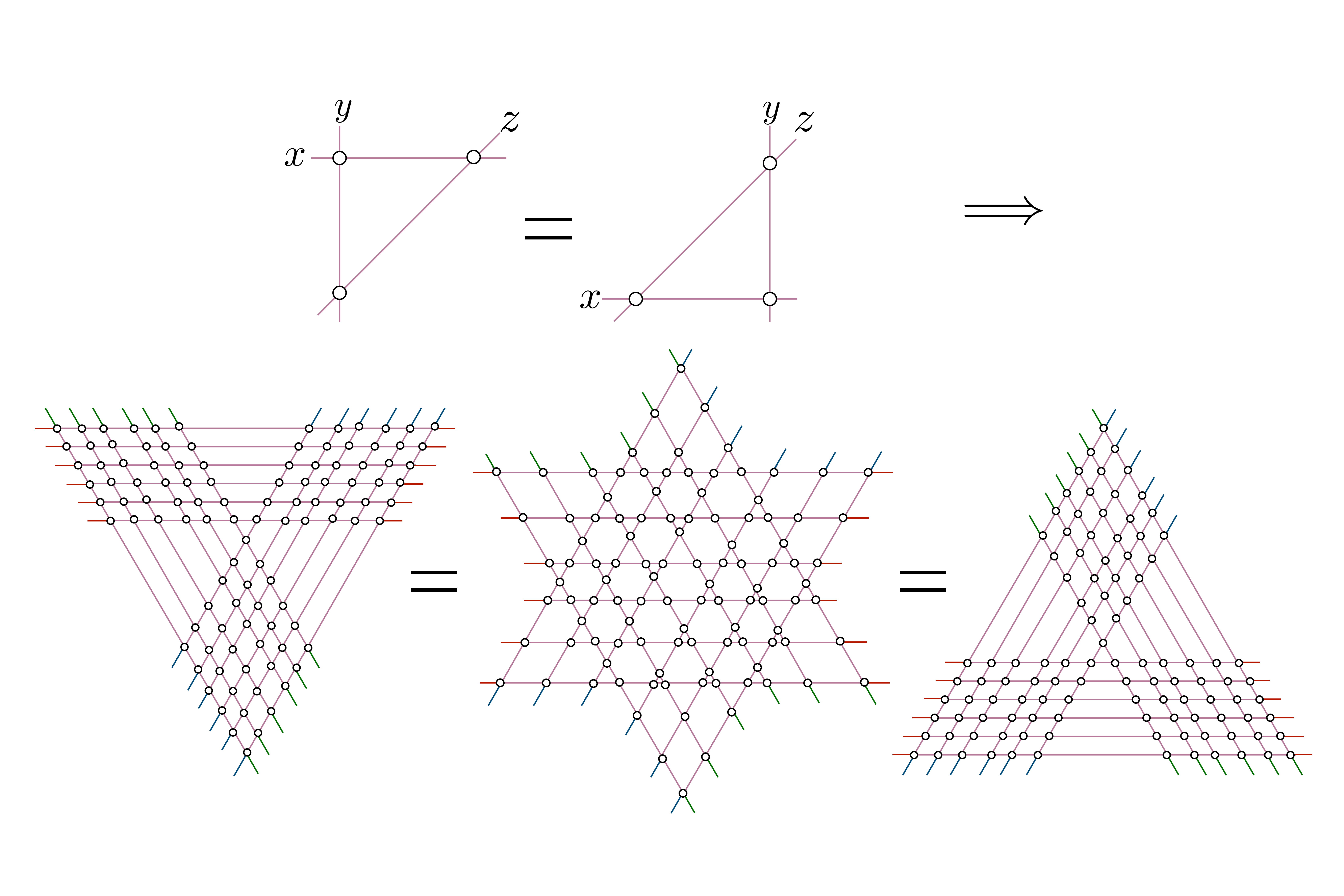}
\caption{The PSU(2|2) vertex satisfies the Yang-Baxter relation illustrated in the first line. Simply put, the partition function is invariant under translating lines across vertices. These allows for many possible rewritings of the partition function, as expressed in the second line. The term on the left-hand side corresponds to the second line of (\ref{HexagonPartition}). The next two terms give alternative representations whose explicit form we hope is clear from the picture. In particular, the middle shape relates the hexagon partition function to a Hubbard-type model in a Kagome-type lattice.}
\label{YangBaxter}
\end{figure}

Alternatively we can think of it as an integrable 1d quantum mechanical model. In this picture we start with three sets of fermions in an in-state and let them scatter -- in a purely factorized fashion -- into a final state also with three sets of fermions. (At intermediate time steps these fermions can scatter into bosons as well but in the initial and final states we only consider fermions.)
These fermions have an $SU(2)$ flavour index and we contract each of the six sets of fermions (three in plus three out) with the same spinor; in other words, we scatter three groups of identical fermions. Because these fermions are identical scattering among the fermions of the same type will be trivial and this is why the big partition function in \ref{triangle}a ends up simplifying to the hexagon shape in figure \ref{triangle}b which gives the name to the partition function. In fact, this simplified shape is far from unique. Since the underlying model is integrable the order by which the particles scatter is imaterial so we can alternatively cast this partition function in a myriad of equivalent ways as illustrated in figure \ref{YangBaxter}.

Each particle has a physical momenta which is conserved in each elastic scattering event; it is thus associated to each line in the figure. We parametrize it by a physical rapidity. Since we have three sets of momenta we will have three sets of rapidities which we label as $v_i, u_j, w_k$. What was before a valance four statistical mechanical weight is in this picture the two-to-two scattering matrix of two of the 1d particles:
\begin{align}
&S_{a b}^{c d}(z,z') \equiv  \vcenter{ \hspace{-0.7cm} \includegraphics[width=0.3\textwidth]{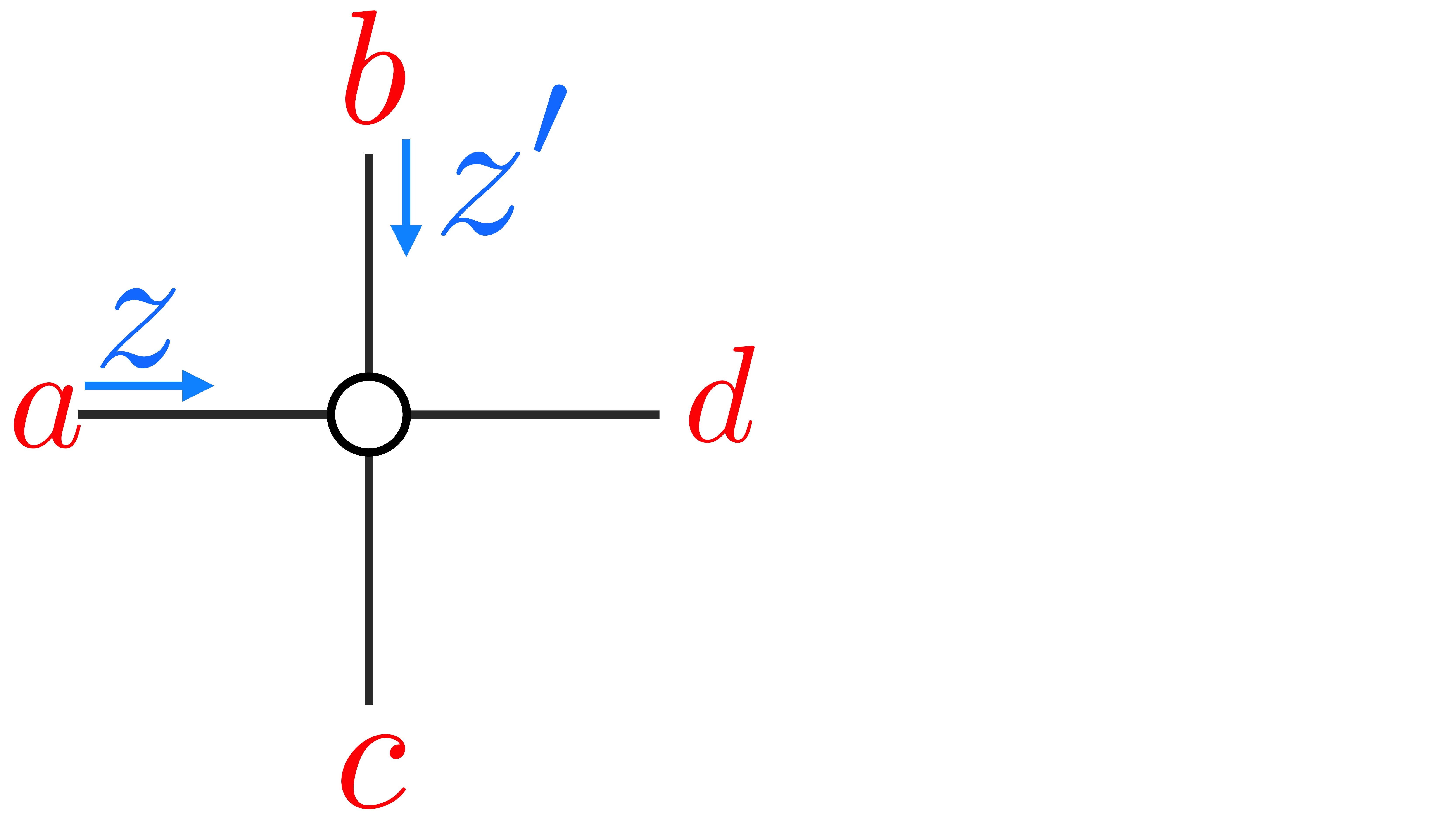}}\hspace{-13.8cm} 
\end{align}
where each index can take four values (a fermion doublet and a boson doublet combined into an $SU(2|2)$ fundamental; for the external states we have fermions only but in intermediate states we will of course produce everyone in the multiplet.)
This S-matrix is Beisert's PSU(2|2) S-matrix \cite{Beisert1, Beisert2} explicitly given by a simple combination of ten terms
\begin{align}
&S_{a b}^{c d}(z,z') =  h(z,z') {\Bigg(}
\mathcal{A}(z,z') 
\frac{\Delta_{a}^{c}\Delta_{b}^{d}+\Delta_{a}^{d}\Delta_{b}^{c}   }{2}   +
\mathcal{B}(z,z') 
\frac{\Delta_{a}^{c}\Delta_{b}^{d} -\Delta_{a}^{d}\Delta_{b}^{c}  }{2}  + \frac{1}{2} \mathcal{C}(z,z') \phi_\mathcal{Z}^{-1}(z,z') E_{a b} \epsilon^{cd}
 \nonumber\\\label{vertex}
&+\mathcal{D}(z,z') \frac{ \delta_{a}^{c}\delta_{b}^{d}+\delta_{a}^{d}\delta_{b}^{c}}{2} 
+\mathcal{E}(z,z') \frac{  \delta_{a}^{c}\delta_{b}^{d}-\delta_{a}^{d}\delta_{b}^{c}}{2}  + \frac{1}{2} \mathcal{F}(z,z') \phi_{\mathcal{Z}}(z,z')\epsilon_{a b} E^{c d}+ \mathcal{G}(z,z') \delta_{b}^{c} \Delta_{a}^{d} \\& + \mathcal{L}(z,z') \delta_{a}^{d} \Delta_{b}^{c} + \mathcal{K}(z,z') \delta_{b}^{d} \Delta_{a}^{c} + \mathcal{H}(z,z') \delta_{a}^{c} \Delta_{b}^{d} \Bigg) . \nonumber 
\end{align}
The Beisert matrix elements $\mathcal{A}, \mathcal{B}, \mathcal{C},  \mathcal{D},  \mathcal{E},  \mathcal{F},  \mathcal{G},  \mathcal{H}, \mathcal{L},  \mathcal{K}$ as well as the non-local markers $\phi_{\mathcal{Z}}$ are defined in appendix \ref{explicit} and depend on the rapidities $z$,$z'$ through the Zhukovsky variables $x^\pm$ only. These are defined through
\beq
\frac{z+i/2}{g} = x^+(z) + \frac{1}{x^+(z)}, \qquad \frac{z-i/2}{g} = x^-(z) + \frac{1}{x^-(z)} \label{Zhukovsky}.
\eeq
where $g = \lambda/(4\pi)^2$ is the coupling. 

The three particle sets are not on equal footing. In our conventions, $u_j$ are \textbf{physical} rapidities while $v_i$ and $w_k$ are \textbf{crossed} kinematics. We can think of the corresponding excitations as anti-particles. Crossed parameters in the matrix elements are to be understood as analytically continued. The result of this analytic continuation is simple in this case: we should pick monodromies around the branch points of the Zhukovsky variables. This amounts to
\beq
x^\pm(v^{\circlearrowright}) \rightarrow 1/x^\pm(v), \qquad x^\pm(w^{\circlearrowleft}) \rightarrow 1/x^\pm(w) \label{ZhuCrossing}
\eeq 
so that when we write, e.g., 
\beq \nonumber \mathcal{A}(z,z') \equiv \frac{x^+(z') - x^-(z)}{x^-(z') - x^+(z)}, \eeq
we are simultaneously defining
\beq
\mathcal{A}(v,w) = \frac{1/x^+(w) - 1/x^-(v)}{1/x^-(w) - 1/x^+(v)}, \qquad \hspace{-0.5cm} \mathcal{A}(v,u) = \frac{x^+(u) - 1/x^-(v)}{x^-(u) - 1/x^+(v)},\qquad \hspace{-0.5cm}\mathcal{A}(u,w) = \frac{1/x^+(w) - x^-(u)}{1/x^-(w) - x^+(u)}\nonumber.
\eeq
All other factors are treated in the same simple fashion; the exception is the 
overall factor $h(x,y)$ given by \cite{hexagons} 
\begin{align}
h(z,z') & \equiv \frac{x^-(z)-x^-(z')}{x^-(z) - x^+(z')}\frac{1-\frac{1}{x^-(z) x^+(z')}}{1-\frac{1}{x^+(z) x^+(z')}}\sigma(z,z')^{-1},
\label{dynFactor}
\end{align}
with $\sigma$ the BES dressing phase \cite{BES} and which transforms nontrivial due to the non-trivial crossing transformation of $\sigma$ under the $v^\circlearrowright$ and $w^\circlearrowleft$ monodromies, see appendix \ref{sigmacrossing}. In any case, this overall factor will lead to an overall factorized product sitting outside as an overall normalization; it is irrelevant for most of our discussion. 

As emphasized above, each line segment in the partition function is a PSU$(2|2)$ state $|x\rangle$ parametrized by two component polarizations $c_\alpha$ and $d_a$ as \beq
\nonumber |x\rangle = \underbrace{c_\alpha |\psi^\alpha\rangle}_{\text{fermions}}  + \underbrace{d_a |\phi^a\rangle}_{\text{bosons}} \equiv (c_1, c_2, d_1, d_2)^T.\eeq In this basis, the matrices $\delta, \Delta, \epsilon, E$ act as 
\beq
\delta = \begin{pmatrix}
1 & 0 & 0 & 0 \\
0 & 1 & 0 & 0 \\
0 & 0 & 0 & 0 \\
0 & 0 & 0 & 0 \\
\end{pmatrix}, \qquad  \hspace{-0.5cm}
\Delta = \begin{pmatrix}
0 & 0 & 0 & 0 \\
0 & 0 & 0 & 0 \\
0 & 0 & 1 & 0 \\
0 & 0 & 0 & 1 \\
\end{pmatrix}, \qquad  \hspace{-0.5cm}
\epsilon = \begin{pmatrix}
0 & 1 & 0 & 0 \\
-1& 0 & 0 & 0 \\
0 & 0 & 0 & 0 \\
0 & 0 & 0 & 0 \\
\end{pmatrix}, \qquad  \hspace{-0.5cm}
E = \begin{pmatrix}
0 & 0 & 0 & 0 \\
0 & 0 & 0 & 0 \\
0 & 0 & 0 & 1 \\
0 & 0 & -1 & 0 \\
\end{pmatrix}. \nonumber
\eeq 
The boundary conditions are so that initial and final states are purely fermionic. Particles in the sets $u, v, w$ have, respectively, incoming polarization spinors $L_1, R_2, R_3$ and outgoing polarization spinors $R_1, L_2, L_3$\footnote{This weird choice of labels is made to simplify formulas in section \ref{SecCtoH}, in which $L_i$ and $R_i$ will correspond to physical left and right space-time polarization spinors of local operators in $\mathcal{N}=4$ SYM. Particle sets $v_i$ and $w_k$, being crossed, have their right and left handed spinning components exchanged. It will also simplify some of the formulas to follow, see e.g. equation (\ref{permutation}).}. In the bulk of the partition function, coloured in pink in figure \ref{triangle}, the particles can transmute into scalars and explore the full PSU$(2|2)$ dynamics.

Having defined all these elements, we can translate picture \ref{triangle} into an equation for the partition function. This is equation (\ref{HexagonPartition}). Note the vertex simplify dramatically for identical fermionic outgoing states, the matrix part being reduced to the element $\mathcal{D} = -1$. This was used to write the last three factors in equation (\ref{HexagonPartition}).  

\begin{align}  \label{HexagonPartition} Z_{J_1,J_2,J_3}  &\equiv  \prod_{i=1}^{J_2} \prod_{j=1}^{J_1}  \prod_{k=1}^{J_3}  \underbrace{{L_1}^{b_j^1}   {R_2}^{a_i^1}   {R_3}^{c_k^1}}_{\text{incoming b.c.}} \times \underbrace{{R_1}_{b_j^{J_2 + J_3}}   {L_2}_{a_i^{J_1 + J_3}}   {L_3}_{c_k^{J_1+J_2}}}_{\text{outgoing b.c.}}\times  \underbrace{(-1)^{J_2 + J_3}}_\text{crossing factor}  \times  \\& {\underbrace{S_{a_{i}^j b_j^i}^{a_i^{j+1} b_j^{i+1}}(v_{J_2+1-i}, u_j)}_{{\color{Orchid}{\text{$v_i$, $u_j$ scattering}}}}}\times \underbrace{S_{a_i^{J_1+k} c_k^i}^{ a_i^{J_1 + k +1} c_k^{i+1}}(v_{J_2 +1 - i}, w_k)}_{{\color{Orchid}{\text{$v_i$, $w_k$ scattering}}}} \times \underbrace{S_{b_j^{J_2+k} c_k^{J_2 + j}}^{ b_j^{J_2 + k +1} c_k^{J_2 + j +1}}(u_{J_1 +1 - j} , w_k)}_{{\color{Orchid}{\text{$u_j$, $w_k$ scattering}}}}  \times \nonumber \\  &  \prod_{i'<i} \underbrace{(-1)^{J_2(J_2-1)/2}h(v_{i'},v_i)}_{{\color{ForestGreen}{\text{$v_{i'}, v_i$ scattering}}}} \prod_{j'<j} \underbrace{(-1)^{J_1(J_1-1)/2} h(u_{j'},u_j)}_{{\color{Maroon}{\text{$u_{j'}, u_j$ scattering}}}} \times \prod_{{k'}<k}  \underbrace{(-1)^{J_3(J_3-1)/2} h(w_{k'},w_k)}_{{\color{NavyBlue}{\text{$w_{k'}, w_k$ scattering}}}} \nonumber .
\end{align}

\subsection{Properties of the partition function}

The $PSU(2|2)$ vertex enjoys four important properties which we will make use in the next few sections. First, it satisfies the Yang-Baxter relation
\begin{equation}
\label{ybeq} S_{b c}^{\alpha \beta}(z',z'') S_{a \alpha}^{d \gamma}(z,z'')S_{\gamma \beta}^{e f}(z,z')= S_{a b}^{\beta \gamma}(z,z') S_{\gamma c}^{\alpha f}(z,z'') S_{\beta \alpha}^{d e}(z',z'')
\end{equation}
which allows for many possible rewritings of the partition formula (\ref{HexagonPartition}) as illustrated in figure~\ref{YangBaxter}. Second, it satisfies the unitarity relation
\begin{equation}
S_{a b}^{\alpha \beta}(z,z') S_{\alpha \beta}^{c d}(z', z) = h(z,z') h(z', z) \label{unitarity},
\end{equation}
as follows from the unitarity of the Beisert PSU$(2|2)$ S-matrix. 

\begin{figure}[t]
\centering
\includegraphics[width=0.9\textwidth]{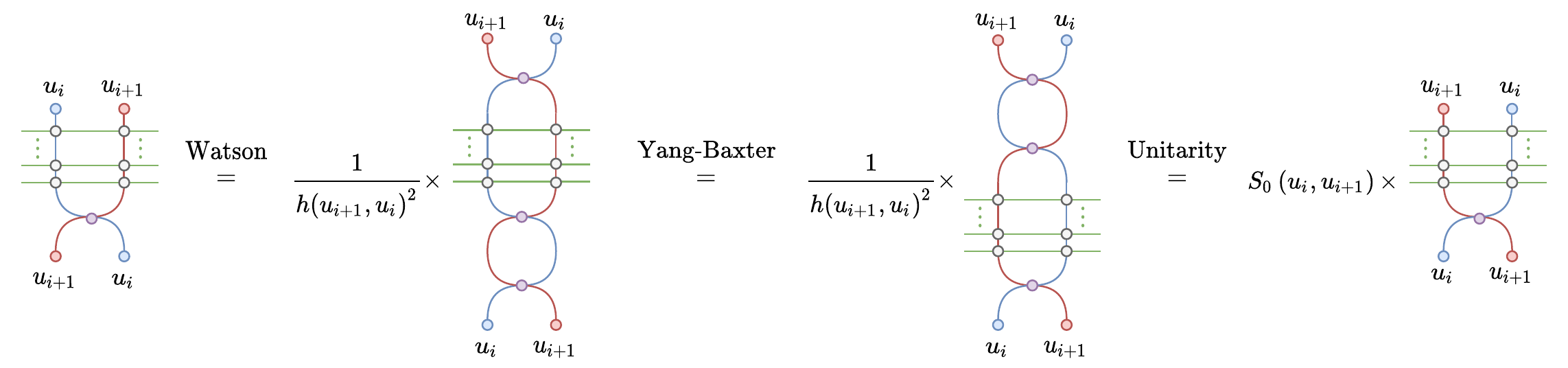}
\caption{The Watson axiom imply that particles in the same external set can be exchanged through the action of the $SL(2,R)$ S-matrix $S_0(x,y) \equiv h(x,y)/h(y,x)$. This follows explicitly from the fact that for identical outgoing polarizations the vertex (\ref{vertex}) projects into the $\mathcal{D}$ element. In equations, ${R_1}_c {R_1}_d S^{c d}_{a b}(u_i, u_{i+1})= -{R_1}_a {R_1}_b h(u_i,u_{i+1})= S_0(u_i,u_{i+1}) {R_1}_c {R_1}_d S^{c d}_{a b}(u_{i+1}, u_{i})$. }
\label{watsonish}
\end{figure}

Note that unitarity immediately allows us to show that, up to a trivial overall factor, the hexagon partition function is an invariant function under swapping of any fermions of the same type as illustrated in figure \ref{watsonish}. More precisely we can decompose any permutation of fermions within the same group as a sequence of neighbouring swaps, each of which simply generates an abelian S-matrix factor $S_0(z_i, z_{i-1}) \equiv h(z_{i},z_{i-1})/h(z_{i-1},z_{i})$,
\begin{equation}
Z_{J_1,J_2,J_3}\Bigr|_{z_i \leftrightarrow z_{i-1}} =  S_0(x_i, x_{i-1})   Z_{J_1,J_2,J_3} \label{symmetry}
\end{equation}
This is the so-called Watson relation.\footnote{Note that we could have redefined the partition function by a trivial product of $h$ factors so that the symmetry relation (\ref{symmetry}) would simplify to full invariance under any swap of rapidities. We found it better \textit{not} to do so to preserve some nice analytic properties. }

The two properties above allow one to move lines around (Yang-Baxter) and collapse bubbles (Unitarity). The next two properties lead to situations which prepare the lines to be moved. 

First, at equal rapidities, the scattering vertex degenerates. For example, when $v_i \rightarrow w_k$\footnote{Here we mean to take $w_k$ and $v_i$ to be equal \textit{after} performing the crossing monodromies $v_i^{\circlearrowright}$ and $w_k^{\circlearrowleft}$.} we have, up to regular terms,
\begin{align}
&S_{a b}^{c d}(v_i,w_k)  \sim \frac{i/\mu(w_k)}{v_i-w_k}\left(-\delta_a^c \delta_b^d -\delta_a^c \Delta_b^d  -\Delta_a^c \delta_b^d  -\Delta_a^c \Delta_b^d\right) =  \vcenter{ \hspace{-0.7cm} \includegraphics[width=0.3\textwidth]{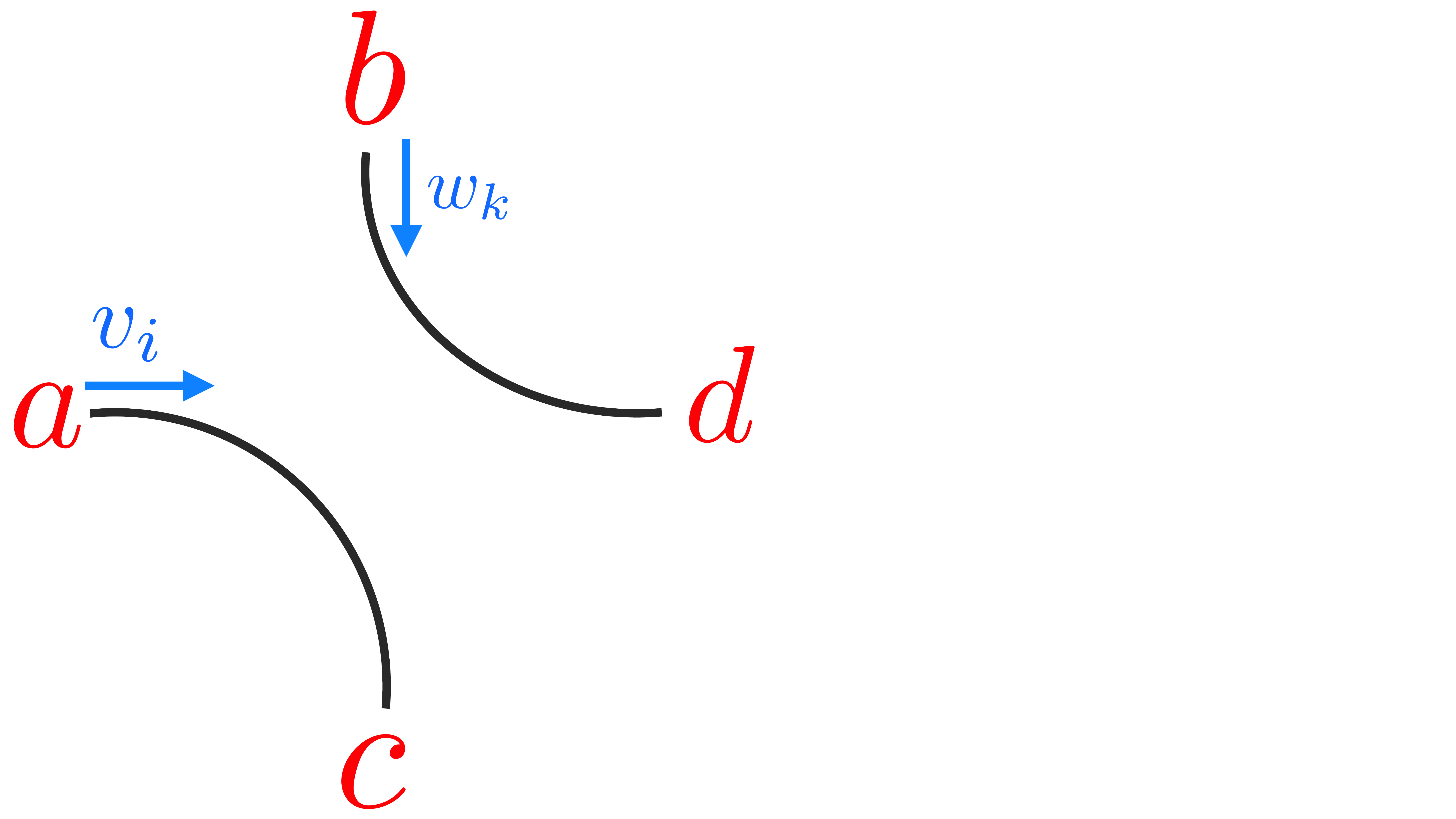}}\hspace{-13.8cm}. \label{degenerates}
\end{align}
In the limit, the vertex develops a pole proportional to the index permutation operator, the residue of which defines the measure $\mu(w_k)$. Physically we can interpret this singularity as a a clustering property (two far away particles disconnect from the rest).  As illustrated in the figure, at this point the two lines meeting at a vertex are now disentangled; they are thus ready to be moved around with unitarity and Yang-baxter in a sequence of moves which can dramatically simplify the partition function. 
\begin{figure}[t]
\centering
\includegraphics[width=1\textwidth]{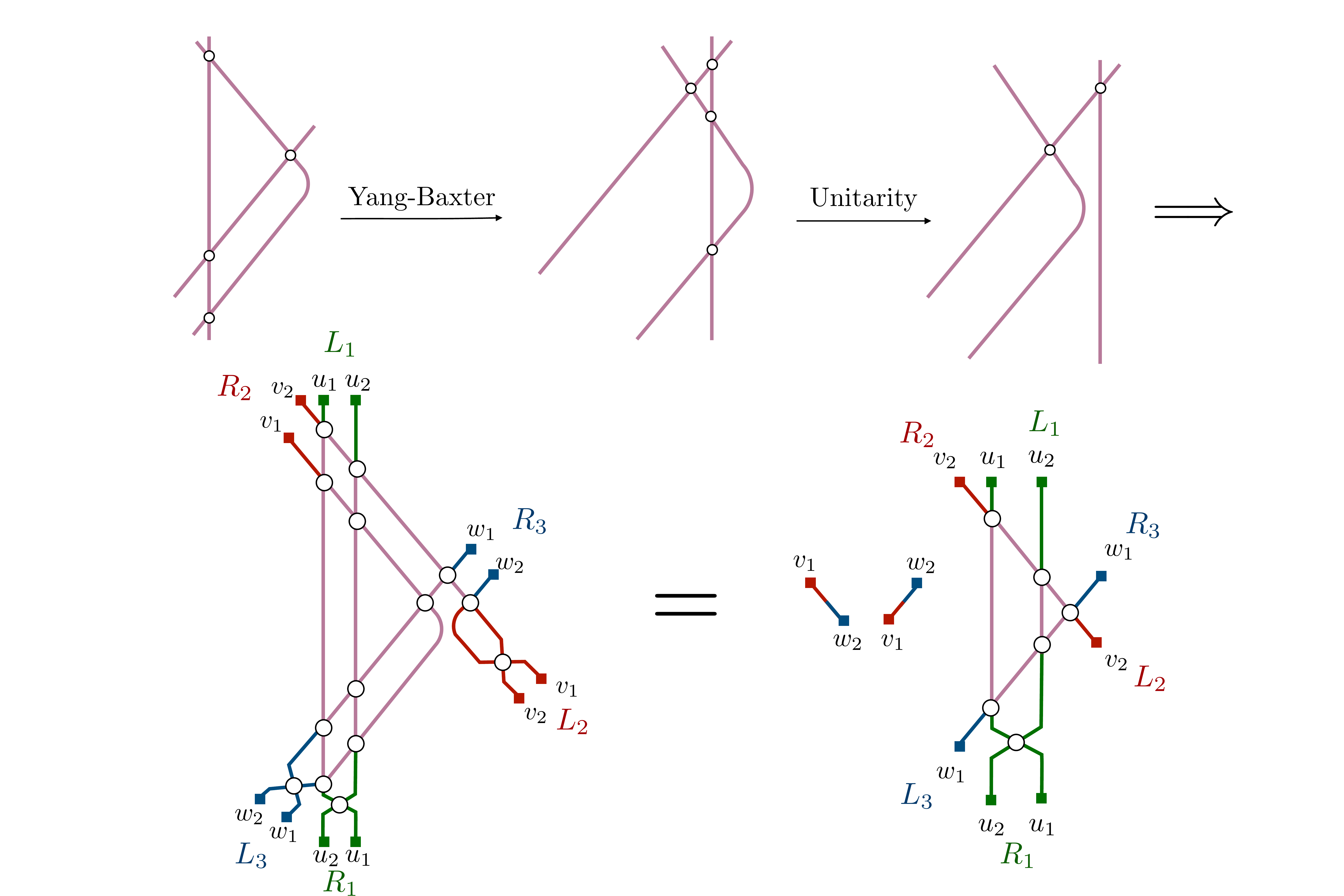}
\vspace{-0.5cm}
\caption{Decoupling limit in the $Z_{2,2,2}$ case. Top: lines can be disentangled through the basic Yang-Baxter and Unitarity moves. Bottom:  When $v_1 \rightarrow w_2$ the vertex degenerate into the permutation operator, see equation (\ref{degenerates}).  One can then use the basic moves (\ref{ybeq}, \ref{unitarity}) to completely factorize the depende on $v_1$ and $w_2$, leading to equation (\ref{decoupling}). In doing so one makes use of the unitarity identities $h(v^{\circlearrowright}_i,v^{\circlearrowright}_j)  h(v^{\circlearrowright}_j,v^{\circlearrowleft}_i) = h(w^{\circlearrowleft}_j,w^{\circlearrowleft}_j)  h(w^{\circlearrowright}_i,w^{\circlearrowleft}_j)  = h(v^{\circlearrowright}_i,u_j)  h(u_j,v^{\circlearrowleft}_i) = 1 , \label{CR-Unity}$.}
\label{decouplingpic}
\end{figure}

\begin{figure}[t]
\centering
\vspace{0cm}
\includegraphics[width=1\textwidth]{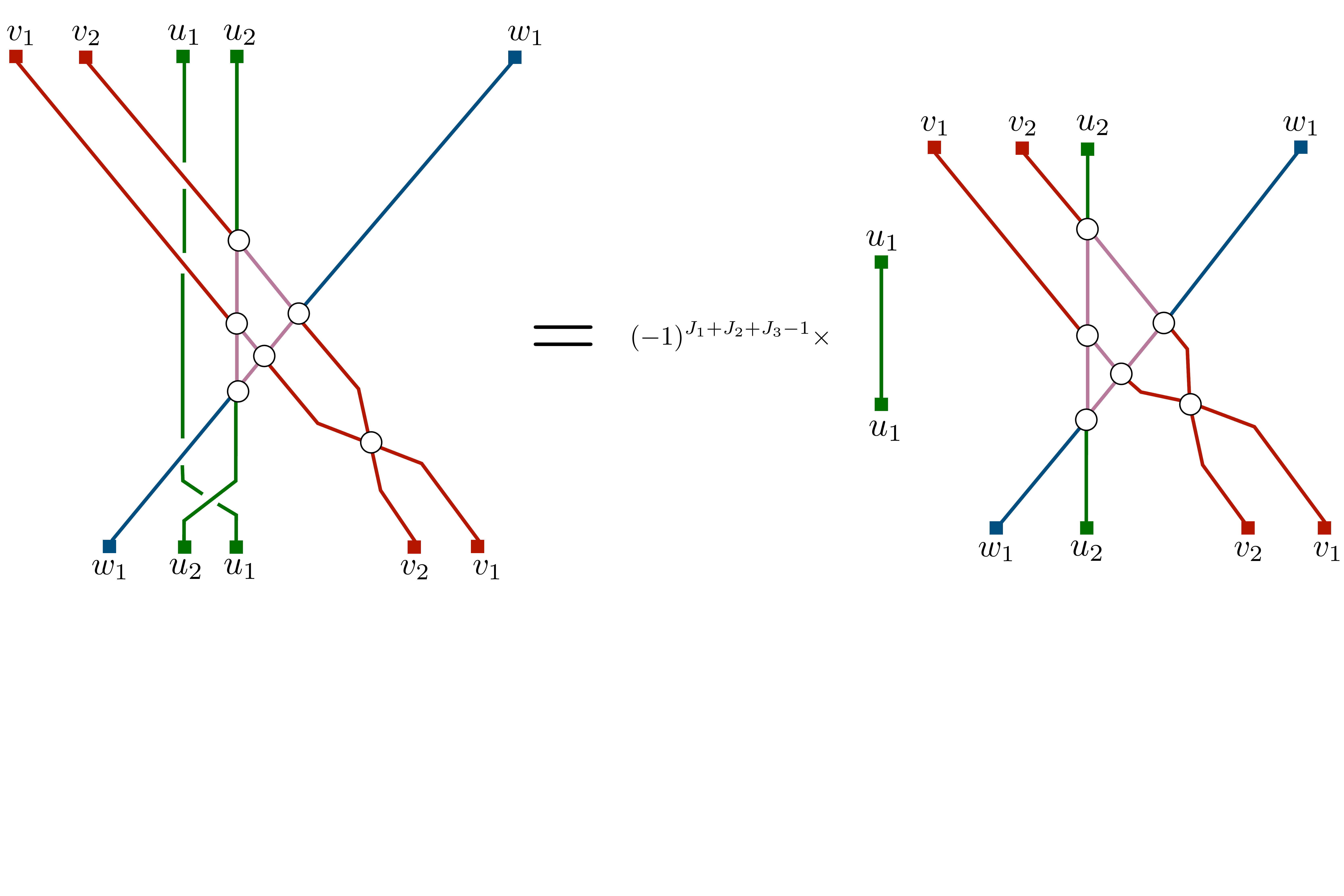}
\vspace{-4cm}
\caption{Decoupling when $u_1 \rightarrow \infty$ for $Z_{2, 2,1}$. When one of the rapidities is sent to infinity the vertex reduces to the free S-matrix. All scatterings of particle $u_1$ are of the fermion-fermion type, resulting in the $(-1)^{J_1 + J_2 + J_3 -1}$ factor of equation (\ref{inftyequation}). }
\label{infinitypartitiondecouple}
\end{figure}

Finally, the vertex also simplifies in the infinite rapidity limit. We have, up to $1/u$ corrections
\begin{equation}
S_{a b}^{c d}(v,u)  = S_{b a}^{d c}(u,w)  \sim \left(-\delta_a^d \delta_b^c +\delta_a^d \Delta_b^c  +\Delta_a^d \delta_b^c  +\Delta_a^d \Delta_b^c\right) +O(1/u)= \vcenter{ \hspace{-0.7cm} \includegraphics[width=0.3\textwidth]{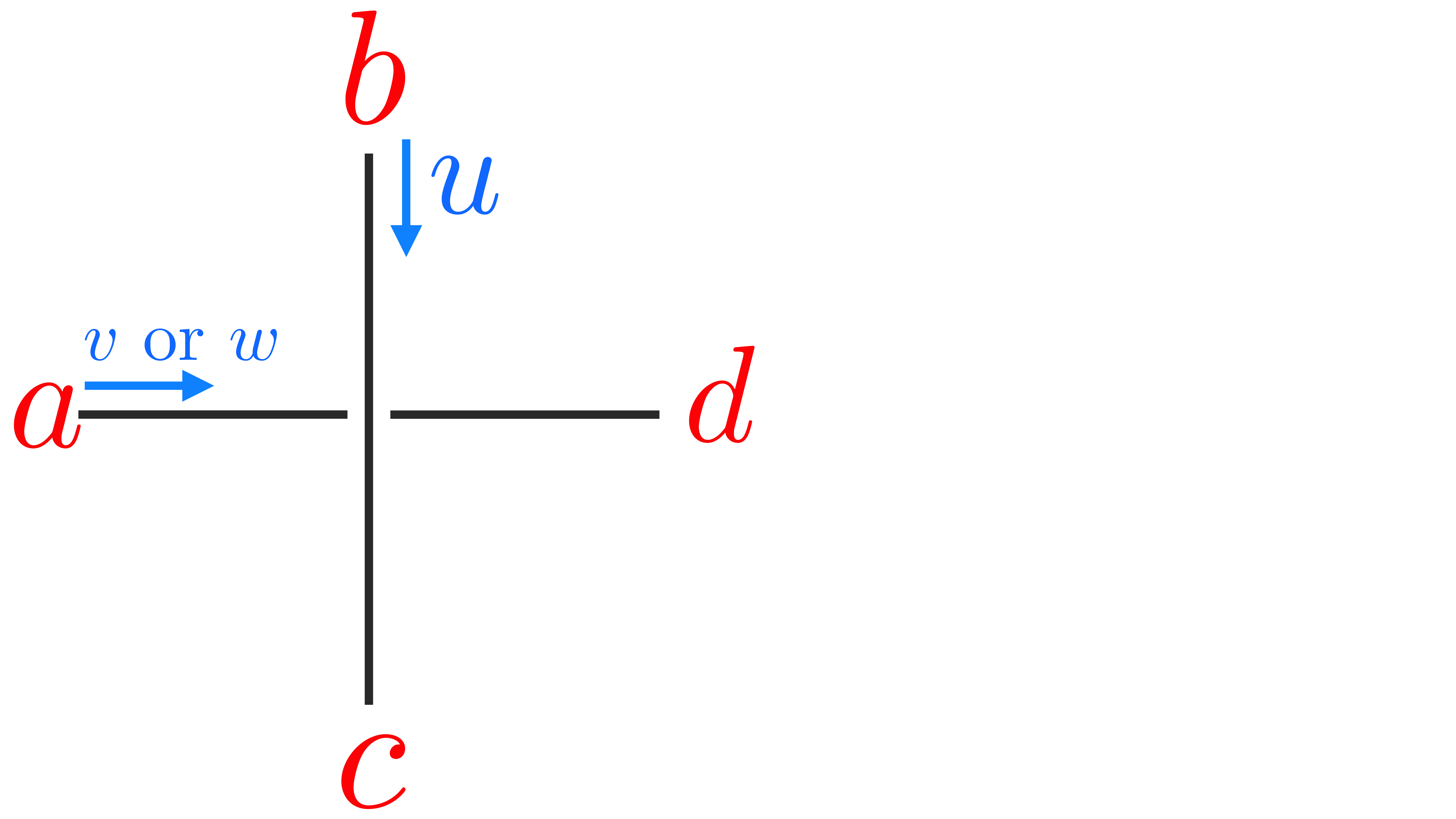}}\hspace{-13.8cm}, \label{infinity} 
\end{equation}
the minus sign in the first term corresponding to the trivial scattering of anticommuting fermions.  Physically, when the rapidity is sent to infinity the particle has zero momenta and became a goldstone excitation which scatters trivially with other particles. After this limit the two lines also became two disentangled pairs, albeit a different choice of pairs compared to the decoupling example. In this case the particle whose root is sent to infinity scatters trivially with all other particles so the simplification of the partition function implies right away that in the $u_1 \rightarrow \infty$ limit, we simply have
                  
\begin{equation}
Z_{J_1,J_2,J_3} \sim (-1)^{J_1 + J_2 + J_3 + 1}  \langle L_1, R_1 \rangle Z_{J_1-1,J_2,J_3}^{u_1}, \label{inftyequation}  \qquad {u_1 \to \infty}
\end{equation} 
which is depicted in figure \ref{infinitypartitiondecouple}. Here and below, rapidities in a superscript in a partition function $Z$ indicates that the rapidities in the superscript are \textit{absent} from the partition function.

These four properties can be used to derive recursion formulas for the partition function. When a vertex degenerates into a permutation operator we can sequentially apply the Yang-baxter and unitarity properties to completely factorize the two decoupling particles from the rest of the partition function, as explained in figure \ref{decouplingpic}.  The result is that the singular term in the $v_1 \rightarrow w_{J_3}$ limit is
\begin{equation}
Z_{J_1,J_2,J_3} \sim - \frac{i}{\mu(w_J)} \frac{\langle L_2, R_3\rangle \langle L_3, R_2 \rangle }{v_1-w_{J_3}} {Z_{J_1,J_2-1,J_3-1}\Bigr|}_{v_a \rightarrow v_{a+1}}, \label{decoupling}
\end{equation}
with $\langle L_a, R_b\rangle \equiv {L_a}_i R_b^i$. This relation was particularly easy to visualize graphically because we decoupled neighboring particles, namely the last particle of type $w$ with the first particle of type $v$, see kets and bras in figure (\ref{triangle}) where that is even more manifest. However, since we can rearrange any particles within any set using (\ref{symmetry}) we can immediately write down the decoupling between any $v$ and any $w$ particle,
\begin{equation}
Z_{J_1,J_2,J_3} \sim -\frac{i}{\mu(w_j)} \frac{\langle L_2, R_3\rangle \langle L_3, R_2 \rangle }{v_i-w_{k}}  \prod_{i'=1}^{i-1} S_0(v_{i'},v_{i}) \prod_{k'=k+1}^{J_3} S_0(w_{k},w_{k'}) {Z_{J_1,J_2-1,J_3-1}\Bigr|}_{v_a \rightarrow v_{a+1}}, \label{decoupling2}
\end{equation}

Decouplings for other pairs $v_i$, $w_k$ can be obtained by moving the particle $v_i$ to the position of $v_1$, and similar for $w_{k}$ and $w_{J_3}$, by means of Watson equation, (\ref{symmetry}) .

\begin{figure}[t]
\centering
\vspace{-0.5cm}
\includegraphics[width=1\textwidth]{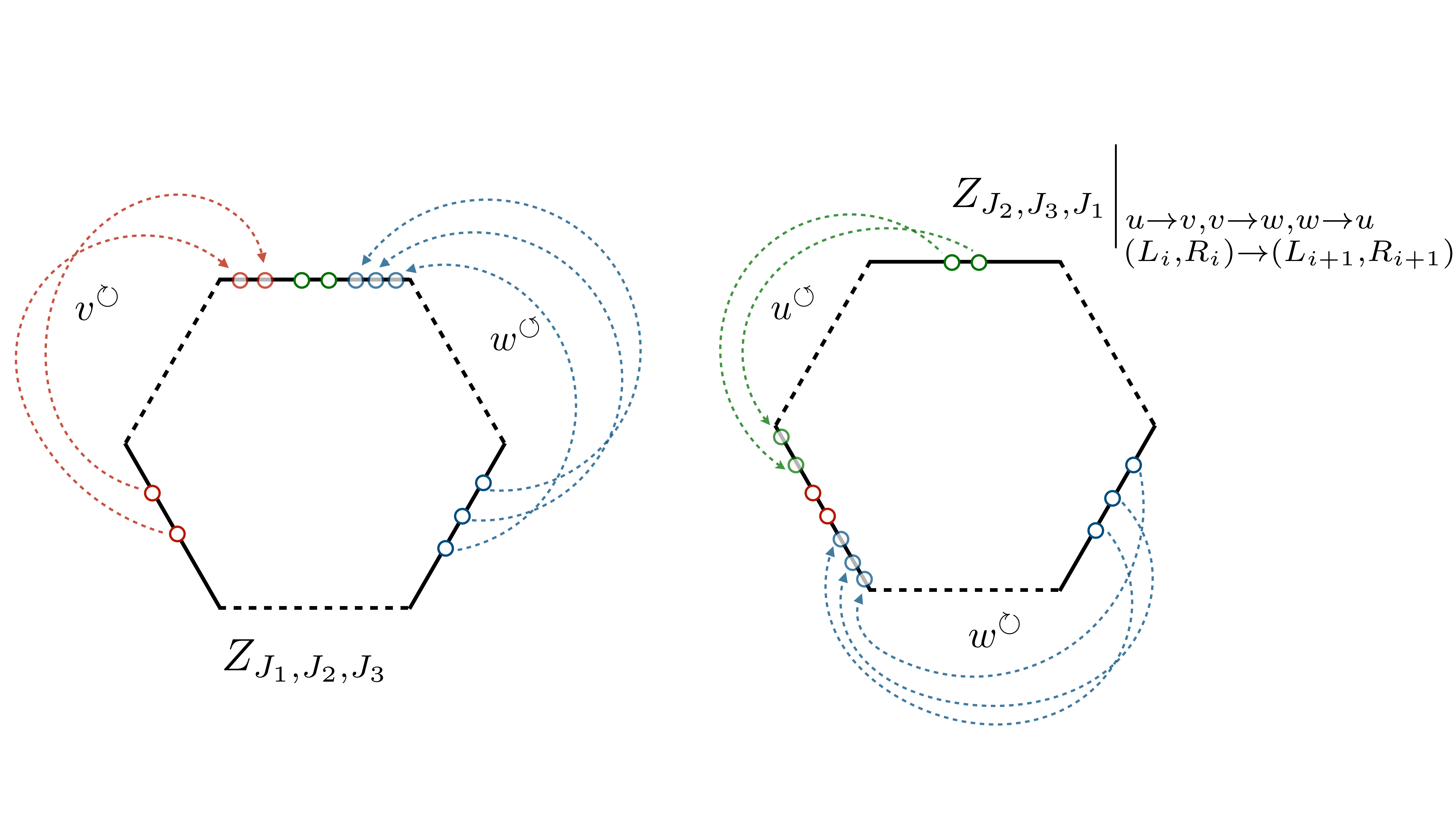}
\vspace{-1.5cm}
\caption{The $\mathcal{N}=4$ hexagon form factor \cite{hexagons} from which the hexagon partition function originates is symmetric in its three physical edges, represented here by solid lines. The expression (\ref{HexagonPartition}) is obtained by moving all particles from the $v$ and $w$ edges to the top $u$ edge, left picture. Moving particles to non-adjacent edges of the hexagon correspond to crossing transformations. Alternatively, the hexagon form factor could be evaluated through the prescription of the right figure, in which $v$ particles remain physical while $w$ is crossed clockwise and $u$ crossed anticlockwise. The result is equation (\ref{permutation}). An equivalent statement is that the hexagon form factor is invariant under crossing all particles in the same direction.}
\label{crossing}
\end{figure}

Crossing symmetry relates different expressions for the partition function that exchange the various sets of particles. We illustrate this in figure \ref{crossing}. In particular, the expression (\ref{triangle}) is invariant under $u\rightarrow v, v \rightarrow w, w \rightarrow u$ with $L_i \rightarrow L_{i+1}, R_i, \rightarrow R_{i+1}$:
\begin{equation}
\qquad \qquad \qquad Z_{J_1,J_2,J_3} = Z_{J_2,J_3,J_1}\bigg\rvert_{\substack{\hspace{-0.65cm} u\rightarrow v, v \rightarrow w, w \rightarrow u\\(L_i,R_i) \rightarrow (L_{i+1},R_{i+1})}}. \label{permutation}
\end{equation}

Combining permutation invariance (\ref{permutation}) with the $v \rightarrow w$ decoupling (\ref{decoupling}) provides $u \rightarrow v$ and $u \rightarrow w$ decoupling formulas 
\begin{align}
Z_{J_1,J_2,J_3} \sim &-\frac{i}{\mu(v_i)} \frac{\langle L_1, R_2\rangle \langle L_2, R_1 \rangle }{u_{j}-v_i} \prod_{j'=1}^{j-1} S_0(u_{j'},u_{j}) \prod_{i'=i+1}^{J_2} S_0(v_{i},v_{i'}) \times   Z_{J_1-1,J_2-1,J_3}^{u_j,v_i}, \label{uvdecoupling}\\
Z_{J_1,J_2,J_3} \sim &\frac{i}{\mu(w_k)} \frac{\langle L_1, R_3\rangle \langle L_3, R_1 \rangle }{u_{j}-w_k} \prod_{j'=j+1}^{J_1} S_0(u_{j},u_{j'}) \prod _{k'=1}^{k-1} S_0(w_{k'},w_{k}) \times   Z_{J_1-1,J_2,J_3-1}^{u_j,w_k}.\label{uwdecoupling}
\end{align}
Once again we write only the singular terms in the $u_j \rightarrow v_i$ and $u_j \rightarrow w_k$ limits respectively.

The equations (\ref{inftyequation}) and (\ref{uvdecoupling}, \ref{uwdecoupling}) in boxes are powerful constraints. They relate partition functions with a certain number of particles (lhs's) with partition functions with less particles (rhs's). An obvious question is whether we can turn them into recursion relations which completely determine the partition function in terms of trivial tiny partition functions with a vertex or two. We could not find a way to do it at finite coupling. The basic obstacle is that the partition function is not a simple rational function of the rapidities whose residues are simple, but instead a multi-sheeted object since the Zhukosky variables naturally take values in tori \cite{tori}. Before crossing cuts, all poles are of the type described in (\ref{inftyequation}, \ref{uvdecoupling}, \ref{uwdecoupling}). However, in the second sheet there are extra poles capturing the fusion of the fundamental fermions into bound states. Correspondingly, the residues are no longer simple decoupled partition functions but instead partition functions with external bound-states. One could hope to bootstrap the complete system of partition functions including all possible external bound-states in a recursive way, but we did not achieve that. On the other hand, in perturbation theory the coupling goes to zero and these tori degenerate and simplify into simple (punctured) spheres. In more pedestrian terms, the square roots in the Zhukosky variables disappear and these become simply proportional to the Bethe roots so that the partition function becomes a nice rational function of the Bethe roots whose poles correspond to simple decouplings.\footnote{Even this is actually not totally trivial because of the non-local $\phi$ Z-markers with its own square roots. It turns out that these square roots can be simplify away.} In that case we can indeed convert the above relations into powerful recursion relations. It is what we will do in section \ref{tree}. After that we will consider a particular example where we can evaluate the full hexagon partition function at any value of the coupling.

\subsection{Tree Level}
\label{tree}
When $g \rightarrow 0$ the Zhukovsky variables (\ref{Zhukovsky}) become rational functions of the rapidities. More surprisingly, the full hexagon partition function becomes rational, the various square roots from the matrix elements (\ref{A-Elem} - \ref{K-Elem}) cancelling out. In this section we use rationality to derive recursion relations that efficiently compute the hexagon partition function for any $J_i$.

Rational functions are completely fixed by their singularities and asymptotics. As a function of $u_1$, $Z(J_1, J_2, J_3)$ has two types of poles both of which with clear physical meaning. The first are poles corresponding to the fusion of two physical particles, $u_1$ and $u_j$, into a bound state. These are the singularities contained in the $h(u_1, u_j)$ factors in (\ref{HexagonPartition}). All other singularities at tree level\footnote{At $g \neq 0$ the physical particles may form bound states with the crossed particles, resulting in new singularities. } are decoupling poles (\ref{uvdecoupling},\ref{uwdecoupling}). We are immediately led to the recursion relation
\begin{align}
\hat{Z}_{J_1,J_2,J_3} =  &(-1)^{J_1 + J_2 + J_3 + 1} \label{recursionrelation} \langle L_1, R_1 \rangle  \hat{Z}_{J_1-1,J_2,J_3}^{u_1} +\\&-\sum_{i=1}^{J_2}   \frac{i}{\mu(v_i)} \frac{\langle L_1, R_2\rangle \langle L_2, R_1 \rangle }{u_{1}-v_i}  \prod_{i'=i+1}^{J_2} S_0(v_{i},v_{i'})  \prod _{j=2}^{J_1} \hat{h}(v_i,u_j)^{-1} \ \times   \hat{Z}_{J_1-1,J_2-1,J_3}^{u_1,v_i}  + \nonumber \\
&+\sum_{k=1}^{J_3}\frac{i}{\mu(w_k)} \frac{\langle L_1, R_3\rangle \langle L_3, R_1 \rangle }{u_{1}-w_k} \prod_{j'=2}^{J_1} S_0(u_1,u_{j'}) \prod _{k'=1}^{k-1} S_0(w_{k'},w_{k}) \prod _{j=2}^{J_1} \hat{h}(w_k,u_j)^{-1} \times   \hat{Z}_{J_1-1,J_2,J_3-1}^{u_1,w_k} \nonumber
\end{align}
with $\hat{Z}_{J_1,J_2,J_3} = {Z}_{J_1,J_2,J_3} \prod_{j<j'} h(u_j,u_j')^{-1} $. Here $\hat{h} = h$ computed \textit{without} performing crossing monodromies in the arguments, since in the decoupling limit $u_j$ approaches $v_i$ and $w_k$ without going around Zhukovsky branch points\footnote{These branch points disappear at tree level, the Zhukovsky variables becoming rational. This is an important point. The crossing transformations do not commute with the perturbative expansion. Crossing can only be performed at finite coupling. In a sense, the simplicity of the tree level result encode some of the finite coupling structure through the many properties derived from crossing and used to bootstrap (\ref{recursionrelation}).}. Explicitly, at tree level
\begin{equation}
\hat{h}(x,y) = \frac{x-y}{x-y-i}, \qquad \mu(x) = 1, \qquad S_0(x,y) =\frac{x-y+i}{x-y-i}.
\end{equation}

Repeated use of (\ref{recursionrelation}) reduces the computation of $Z_{J_1,J_2,J_3}$ to that of ${Z}_{0, J_2, J_3} $. One could then use rationality in $v_i$ and $w_k$ to write further recursions that reduce the formula all the way to $Z_{0,0,0} = 1$. One can skip this step, in practice, since the recursions for ${Z}_{0, J_2, J_3} $ can be solved explicitly. Through induction one finds the determinant formula
\begin{equation}
{Z}_{0, J_2, J_3} = (-1)^{A_{J_2 J_3}} \frac{\prod_{i=1}^{J_2}\prod_{k=1}^{J_3}(v_i - w_k + i)}{(\prod_{i<i'} v_{i'} - v_{i} + i)(\prod_{k<k'} w_{k} - w_{k'} - i)} \label{det} \det\text{M}, \nonumber
\end{equation}
where M is a square matrix of size $\max(J_2,J_3)$ whose elements are given by
\begin{equation} 
\hspace{-1cm} M_{i j} =  \begin{cases}
\frac{\langle 23 \rangle \langle 32 \rangle}{v_i - w_j}+\frac{\langle 22 \rangle \langle 33 \rangle}{v_i - w_j+i}-\frac{\langle 23 \rangle \langle 32 \rangle}{v_i - w_j+i}\,, &  \text{ if } i \leq J_2 \land j \leq J_3 \\
 & \\ 
-\frac{\langle 23 \rangle \langle 32 \rangle}{\langle 33 \rangle v_i^{J_3-j+1}}- \frac{\langle 22 \rangle}{(v_i+i)^{J_3-j+1}}+ \frac{\langle 23 \rangle \langle 32 \rangle}{\langle 33 \rangle (v_i+i)^{J_3-j+1}}\,, &\text{ if $j > J_3$}\\
 & \\ 
\frac{\langle 23 \rangle \langle 32 \rangle}{\langle 22 \rangle w_j^{J_2-i+1}}+\frac{\langle 33 \rangle}{(w_j-i)^{J_2-i+1}}-\frac{\langle 23 \rangle \langle 32 \rangle}{\langle 22 \rangle (w_j-i)^{J_2-i+1}}\,,  &\text{ if $i > J_2$}
\end{cases} \nonumber,
\end{equation}
with $\langle i j \rangle \equiv \langle L_i, R_j \rangle $ and
\begin{equation}
 A_{J_2 J_3} = 
(J_2 +1)(J_3+1) + \sum_{i=2}^3(J_i-1)J_i/2  + \begin{cases} 
\frac{(J_3 - J_2)(1 + J_3 - J_2)}{2}\,, & \text{if $J_2<J_3$} \\ 
0\,, & \text{otherwise} 
\end{cases} \nonumber .
\end{equation}

In section \ref{perturbativechecks} the recursive formulas (\ref{recursionrelation} ,\ref{det}) will be used to produce three point structures data for operators of arbitrary twist at tree level. For convenience, we attach a ready to copy-and-paste Mathematica implementation of the tree level recursion. The SL$(2)$ S-matrix and dynamical factor are
\vspace{0.5cm}
{\footnotesize
\verb"          "\\
\verb"  S0[z_, zp_] = (z - zp + I)/(z - zp - I);"\\
\verb"  h[z_, zp_] = (z - zp)/(z - zp - I);"
\verb"          "\\
}\\
We then write the recursive formulas for the partition function. As explained above, when applying the recursion to one of the three particle sets, it is useful to factor out the dynamical factors. We denote this procedure through ``hats" which must be removed to recover $Z$. First we do recursions for ``v's" starting with the trivial one-set answer:

{\footnotesize
\verb"          "\\
\verb" Zhathat[0,J3_] := V[3]^J3 Product[(-1) h[w[k], w[kk]], {kk, J3}, {k, kk - 1}];"\\
\verb" Zhathat[J2_,J3_]:= Zhathat[J2,J3] = (-1)^(J2+J3+1) V[2] (Zhathat[J2-1,J3] /.v[i_]:>v[i+1])-"
\verb"     Sum[I H[2,3]/(v[1]-w[k])Product[S0[w[k],w[kk]],{kk,k+1,J3}] Product[h[w[k],v[i]]^-1,{i,2,J2}]"\\
\verb"      (Zhathat[J2-1,J3-1] /.{v[i_]:>v[i+1],w[kkk_]/;kkk>=k:>w[kkk+1]}),{k,J3}];"\\
\verb" Zhat[0,J2_,J3_]:=Zhathat[J2, J3]Product[h[v[i], v[ii]],{ii,J2},{i,ii-1}];"}

Next, we perform recursions for the ``u" particles to recover the full partition function,

{\footnotesize
\verb"          "\\
\verb" Zhat[J1_,J2_,J3_]:= Zhat[J1,J2,J3]=(-1)^(J1+J2+J3+1)V[1](Zhat[J1-1,J2,J3] /.u[j_]:>u[j+1])-"
\verb"      Sum[I H[1,2]/(u[1]-v[i])Product[S0[v[i],v[ii]],{ii,i+1,J2}]Product[h[v[i],u[j]]^-1,{j,2,J1}]"
\verb"        (Zhat[J1-1,J2-1,J3]/.{u[j_]:>u[j+1],v[iii_]/;iii>=i:>v[iii+1]}),{i,J2}]+"\\
\verb"           Sum[I H[3,1]/(u[1]-w[k])Product[S0[u[1],u[jj]],{jj,2,J1}]Product[S0[w[kk],w[k]], "\\
\verb"             {kk,1,k-1}] Product[h[w[k],u[j]]^-1,{j,2,J1}](Zhat[J1-1, J2,J3-1]/.{u[j_]:>u[j+1],"\\
\verb"              w[kkk_]/;kkk>=k:>w[kkk+1]}),{k,J3}];"\\
\verb"Z[J1_,J2_,J3_]:=Zhat[J1,J2,J3]Product[h[u[j],u[jj]],{jj,J1},{j,jj-1}];"\\
}\\
as promised. In the code we denoted $V[i]\equiv \langle L_i, R_i \rangle$ and $H[i,j]\equiv \langle L_i, R_j \rangle \langle R_i, L_j \rangle$.

\subsection{All Loop Abelian Simplifications}
\label{abelianpartition}
\begin{figure}[t]
\centering
\vspace{0cm}
\includegraphics[width=1\textwidth]{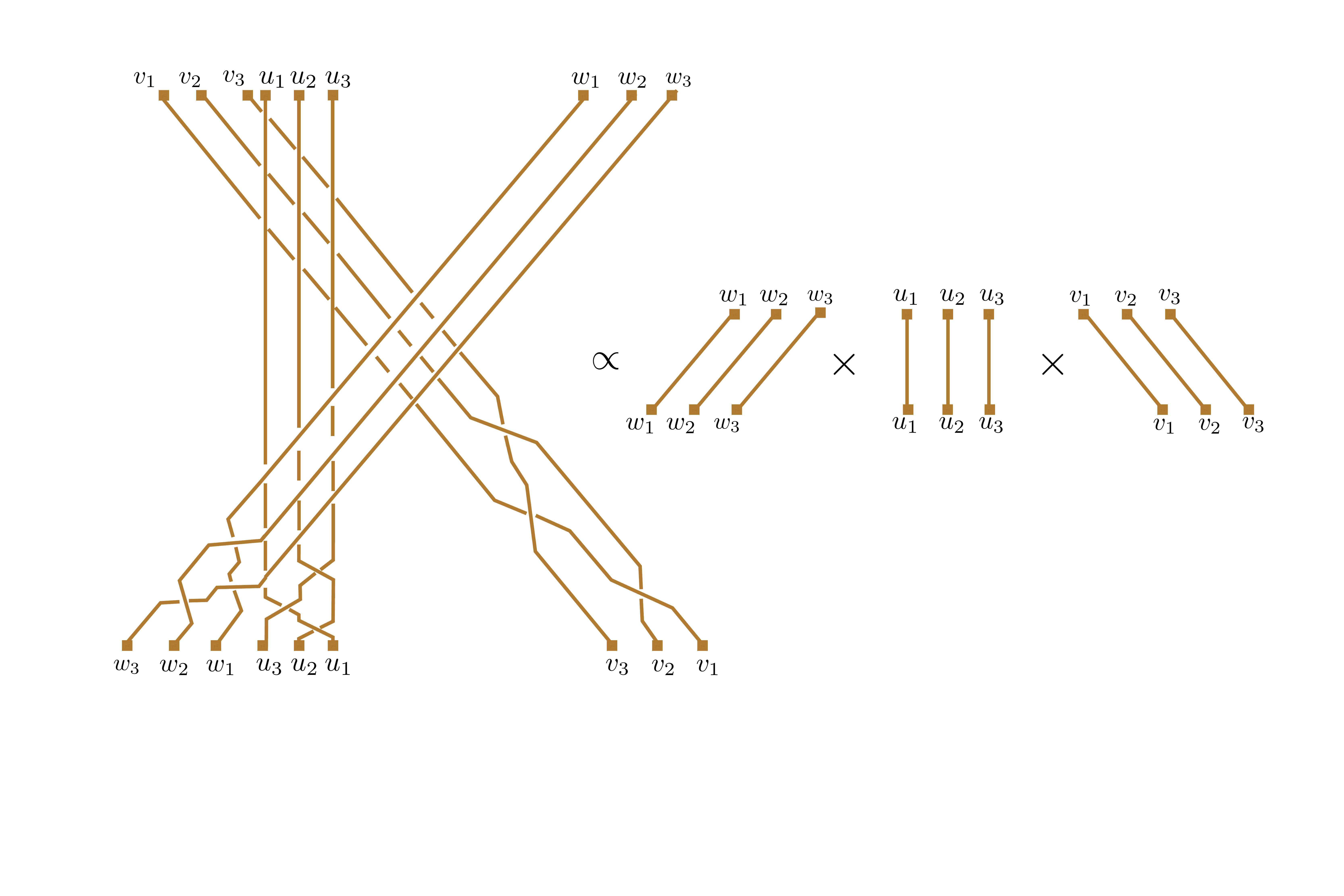}
\vspace{-2.5cm}
\caption{When the incoming state are fermions with parallel polarizations, the vertex becomes proportional to the identity, e.g. ${R_2}^a {L_1}^{b} S^{c d}_{a b}(u_i, u_{i+1})= -{R_2}^c {L_1}^d h(v_2, u_1 )$. This simplification happens at all vertices if $L_1 = R_2 = R_3$, the partition function reducing to the product of abelian factors times the inner product between incoming and outgoing polarizations, equation (\ref{abelian}). Of course, the same simplification holds if instead we have $R_1 = L_2 = L_3$.}
\label{abelianhex}
\end{figure}

When $L_1 = R_2 =R_3$ the triangle partition function dramatically simplifies. In that case, only the terms proportional to the $\mathcal{D}$ matrix element in (\ref{vertex}) survives. As shown in figure \ref{abelianhex}, the result is
\begin{align}
Z^{\text{Abelian}}_{J_1,J_2,J_3} =&  \langle L_1, R_1 \rangle^{J_1}  \langle L_2, R_2 \rangle^{J_2}   \langle L_3, R_3 \rangle^{J_3}  \times (-1)^{J_2 + J_3 + J(J-1)/2}\label{abelian}
 \\ & \prod_{i=1}^{J_2} \prod_{j=1}^{J_1}  \prod_{k=1}^{J_3}  h(v_i, u_j) h(u_j, w_k) h(v_i,w_k)\prod_{\substack{{i'<i}\\{j'<j}\\{k'<k}}}  h(v_{i'},v_i)  h(u_{j'},u_j) h(w_{k'},w_k),\nonumber\end{align}
 with $J = J_1 + J_2 + J_3$. We emphasize this result holds at finite coupling $g$.

 Of course, from permutation invariance, the result also simplifies for $L_2= R_1 =R_3$ and $L_3 = R_1 =R_2$.  This is non-trivial from the expression (\ref{HexagonPartition}) but can be made manifest through crossing transformations.  The vertices are also projected into  the $\mathcal{D}$ element if $R_1 = L_2 =L_3$ or permutations, as can be seen from (\ref{vertex}). The partition function for these polarizations also reduces to formula (\ref{abelian}).  

\section{Spinning hexagons}
\label{SecCtoH} 

\subsection{Polarizing the Hexagon OPE}
Now we shift gears by focusing on the computation of three point functions of primary operators in the SL$(2|\mathbb{R})$ sector of $\mathcal{N}=4$ SYM. One important ingredient in our analysis, as we shall see below, is the hexagon partition function analysed in the previous section. Before delving into those details let us define the precise form of the operators that are being considered and summarize how the hexagon formalism can be used to compute three point functions.

The primary operators in the SL$(2|\mathbb{R})$ sector are traceless symmetric single traces made out of covariant derivative excitations $\mathcal{D}^{\alpha \dot{\alpha}}$ on top of a BPS ``vaccum" of fundamental scalars, $\text{Tr}(\vec{Y} \cdot\vec{\phi})^\tau$. To denote operators of spin $J_i$ we use the index free notation in terms of polarization spinors
\begin{equation}
\mathcal{O}_{J_i}(x) \equiv
\mathcal{O}_{J_i}(x,L_i, R_i) = \left(\prod_{j=1}^{J_i} {L_i}_{\alpha_j}{R_i}_{\dot{\alpha}_j}\right) \mathcal{O}_{J_i}^{\alpha_1 \dots \alpha_{J_i} \dot{\alpha}_1 \dots \dot{\alpha_{J_i}}}(x).
\end{equation}
These operators are the same type of spinning operators we consider in the conformal bootstrap \cite{paper1}, being the only difference here that we do not need to consider leading twist, in fact, here they can have any twist $\tau$, which we leave implicit.

By conformal symmetry, the full three point function can be recovered from its values in the conformal frame\footnote{More generally, by superconformal symmetry, we can also fix the R-symmetry polarizations $Y_i$ to point at fixed directions in $SO(6)$. The dependence on them is completely fixed by the Ward identities. In practice we use the twisted translated polarizations of \cite{hexagons}. The R-symmetry factors will not be important in our discussion and we mostly omit them.}. We place operators along the $x_2$ axis at $0$, $1$ and $\infty$. The rescaled correlator 
\begin{equation}
C(J_1,J_2,J_3)  \equiv \lim_{\mathcal{L} \rightarrow \infty} \mathcal{L}^{2 \Delta_3} \langle \mathcal{O}_{J_1}(0)\rangle
\mathcal{O}_{J_2}(1)
\mathcal{O}_{J_3}(\mathcal{L})
\end{equation}
can then be expressed as a sum over tensor structures invariant under the residual unfixed symmetry \cite{Costa:2011mg, Kravchuk:2016qvl}. As defined in appendix \ref{spinors}, these tensor structures can be written as inner products of the polarization spinors $L_i$ and $R_i$, as follows
\begin{equation}
C(J_1,J_2,J_3) = \sum_{\ell_1,\ell_2,\ell_3} C^{J_1,J_2,J_3}_{\ell_1,\ell_2,\ell_3} \frac{\langle 1,1\rangle^{J_1-\ell_2-\ell_3}\langle 2,2\rangle^{J_2-\ell_1-\ell_3}\langle 3,3\rangle^{J_3-\ell_1-\ell_2}}{\langle 2, 3 \rangle^ {-\ell_1}\langle 3,2\rangle^{-\ell_1}\langle 1, 3 \rangle^{-\ell_2} \langle 3,1\rangle^{-\ell_2}\langle 1, 2 \rangle^{-\ell_3}\langle 2,1\rangle^{-\ell_3}}.
   \label{CinConfFrame}
\end{equation}
where $\langle i,j\rangle = \langle L_i, R_j \rangle$. The goal is to reproduce this three point function -- which at once captures all structure constants for any tensor structures -- from the hexagon formalism. 

\begin{figure}
\centering
\includegraphics[width=0.9\textwidth]{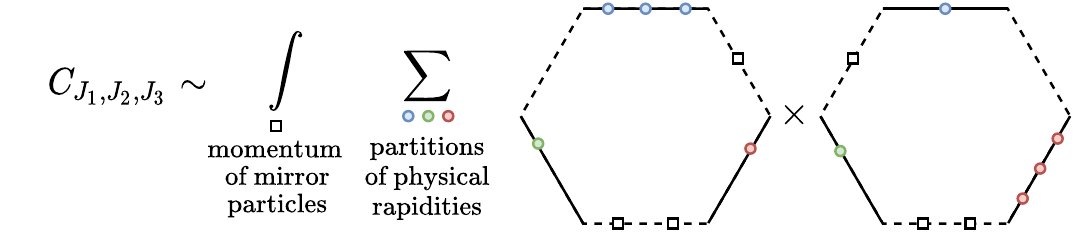}
\caption{Each closed spin chain operator is split into two open chain operators. We sum over all the ways its excitations can end up in either one of the chains. Gluing the hexagons together amounts to integrating over all possible mirror states.}
\label{addHexagons}
\end{figure}

The hexagon formalism has already been developed and tested thoroughly for spinless operators to very high perturbative orders (as well as for spinning correlators without non-trivial tensor structures) so the main novelty here is how to deal with the various spinor polarizations leading to the various terms summed in (\ref{CinConfFrame}). Before presenting the relevant expressions with these spinor boundary conditions accounted for let us recall that the hexagon formalism entails two main components as illustrated in figure \ref{addHexagons}:
\begin{itemize}
\item One is a sum over partitions of physical rapidities for each external operator. When we cut the three point function pair of pants into two hexagons the excitations on each operator can end up on either hexagon so we must sum over ways of partitioning such excitations. These sums are often referred to as the asymptotic sums. 
\item The other are integrals over mirror rapidities along the three seams of the pair of pants. When we glue back the hexagons into a pair of pants we must sum over all possible quantum states along each edge where we glue. These mirror integrals are often referred to as wrapping effects. 
\end{itemize}

\textbf{In this work we completely ignore the second effect}. Except for section \ref{CtoH2loops}.

There are two simple reasons for doing so. The first is that wrapping corrections are delayed in perturbation theory. At tree level and one loop they do not show up even for the smallest twist two external operators for instance and for larger twist operators they are delayed even more in perturbation theory. So we better clean up the asymptotic part first -- where all the subtleties related to the spinors already shows up -- before addressing these finite size mirror corrections. The second reason is that even at loop level, sometimes we can drop the wrapping corrections altogether if the distances travelled by the mirror particles are very large. This happens for large operators with lots of R-charge but also for operators of small R-charge and very large spin. The intuition is that when the spin is very large the centripetal force effectively opens up the operator making it effectively very large; in this case sometimes these finite size corrections can be ignored. We thus suspect this to be the case for the case of three twist two operators with very large spin and generic polarizations. Evaluating this kind of correlator should beautifully connect to the remarkable Wilson loop dualities \cite{paper1,Korchemsky:1992xv,Alday:2010ku}. So if our goal is to eventually prove these dualities from integrability perhaps we can ignore wrapping for now. 

With wrapping out of the window, we can then present our main formula for the spinning three point function. It is a long formula but fortunately most ingredients are self-explanatory and have appeared before in several spinless studies. It reads
\begin{equation}
C(J_1,J_2,J_3) = 
\underbrace{\mathcal{N}(\textbf{u})\mathcal{N}(\textbf{v})\mathcal{N}(\textbf{w})}_{\texttt{normalizations } }
\times \sum_{\substack{{a\, \cup \, \bar{a}=\textbf{u}}\\{b\, \cup \, \bar{b}=\textbf{v}}\\{c\, \cup \, \bar{c}=\textbf{w}}}} 
\underbrace{ \omega_{\ell_{13}}(a,\bar{a})\omega_{\ell_{12}}(b,\bar{b})\omega_{\ell_{23}}(c,\bar{c}) }_{ \texttt{splitting factors}} \times
\underbrace{\mathcal{H}(a,b,c) \times \mathcal{H}(\bar{a},{\bar{c}},{\bar{b}})}_{\texttt{hexagons}}
\label{CtoH}
\end{equation} 
where the normalization and splitting factors are given by
\begin{eqnarray}
\mathcal{N}(\mathbf{u})^2 = \frac{\prod\limits\limits_{i=1}^{J} \mu(u_i)}{\prod\limits_{i\neq j}^{J} S(u_i,u_j) \det{ \left(\partial_{u_i}\phi_j\right)}} \,, \qquad  \omega_{\ell}(a,\bar{a}) = \prod\limits_{u_j \in \bar{a}} (-e^{i p(u_j)\ell})\prod\limits_{\substack{{u_i \in a}\\{i>j}}} S(u_i,u_j)  \,.
\end{eqnarray}
These are the same factors arising already in \cite{hexagons}; the notation and physical meaning of these factors is the same as there.  

\begin{figure}[t!]
\hspace{-8cm}
\includegraphics[width=1.45\textwidth]{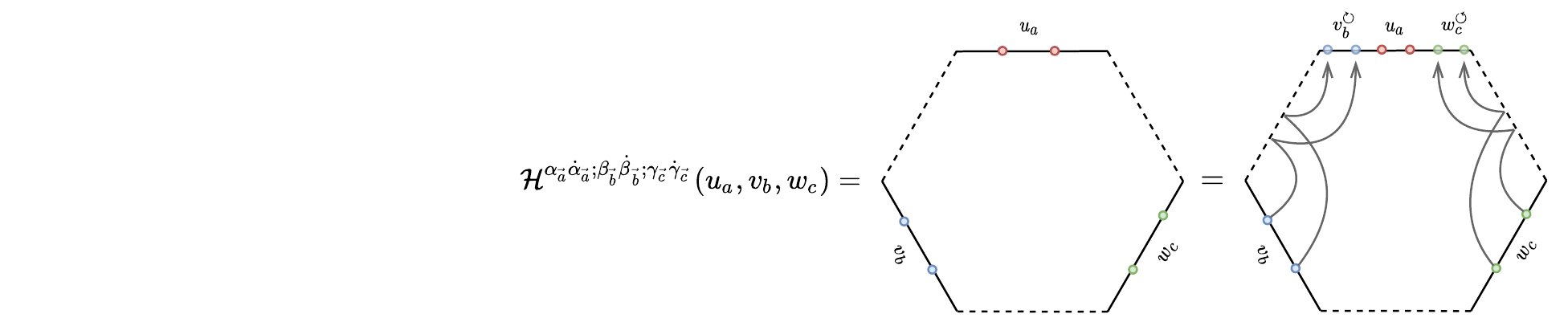}
\caption{To obtain a correlator with all excitations on top we perform one crossing transformation in the left ($\circlearrowright$) and another in the right ($\circlearrowleft$). These monodromies will change the Zhukovsky variables as well as the indices of the excitations.}
\label{HCrossing}
\end{figure}

The most important objects in (\ref{CtoH}) -- and where important novelty lies -- are the hexagons form factors $\mathcal{H}$. We can relate either of them to a purely creation amplitude with all excitations in the top by applying a sequence of crossing transformations, see figure \ref{HCrossing}. This leads to a representation of the hexagon as 

\begin{align}
\mathcal{H}(a,b,c) & = (-1)^\frac{(a+b+c)(a+b+c-1)}{2}\prod_{\substack{{u_i \in a}\\{v_j \in b}\\{w_k \in c}}} 
L_{1,\alpha_1}\dots L_{2,\alpha_{2,j}} L_{3,\alpha_{3,k}} R_{1,\beta_{1,i}}R_{2,\beta_{2,j}}R_{3,\beta_{3,k}}\times \nonumber \\
& \times \langle \mathfrak{h}|\mathcal{D}^{\alpha_{2,j}\beta_{2,j}}(v_j^{\circlearrowright})\mathcal{D}^{\alpha_{1,i}\beta_{1,i}}(u_i)\mathcal{D}^{\alpha_{3,k}\beta_{3,k}}(w_k^{\circlearrowleft})\rangle
\label{CreationAmplitude}
\end{align}
where the minus factor in front is the grading associated with the fermionic nature of the excitations.

This creation amplitude describes the scattering of three sets of fermions, labelled by their rapidities   $\textbf{u}$,  $\textbf{v}$ and  $\textbf{w}$ in the $\mathcal{N} =4$ SYM spin-chain. Each set starts polarized in a fixed direction labeled by their correspondent left-hand spinors. The particles then scatter in all possible pairings according to Beisert's $PSU(2|2)$ vertex, with the final state being the projection into fermions of definite polarization defined by its right-hand spinors. The object that accounts for all these scatterings is the \textbf{hexagon partition function} ($Z$), defined as
\begin{equation}
\mathcal{H}(u_a,v_b,w_c) = (-1)^\frac{(a+b+c)(a+b+c-1)}{2} Z_{a,b,c}\,. \label{HtoZ}
\end{equation}

The particles entering this partition function are not on equal footing, since we used crossing transformations for the rapidities $\textbf{v}$ and $\textbf{w}$. The first effect of these transformations is the exchange their left and right $PSU(2|2)$ indices
\begin{equation}
\mathcal{D}^{\alpha \beta} \xrightarrow{\circlearrowleft} -\mathcal{D}^{\beta \alpha} \;\; \text{and} \;\; \mathcal{D}^{\alpha \beta} \xrightarrow{\circlearrowright} -\mathcal{D}^{\beta \alpha}\,,
\end{equation}
this will swap incoming and outgoing indices for particles $\textbf{v}$ and $\textbf{w}$. More precisely, it will set $L_1$, $R_2$, $R_3$ to be the incoming boundary conditions and $R_1$, $L_2$, $L_3$ to be the outgoing boundary conditions for the scattering, see figure \ref{triangle}a. 

Another implication of crossing is the introduction of \textbf{crossed} parameters in the S-matrix elements. This amounts to analytically continue these matrix elements when considering the crossed rapidities $\textbf{v}$ and $\textbf{w}$. The effects of these analytical continuations are simple: we must pick monodromies around the branch point of the Zhukovsky variables (\ref{ZhuCrossing}) and mind the non-trivial factors introduced by the BES dressing phase, see appendix \ref{sigmacrossing}.

In the next sections we use the properties of the \textbf{hexagon partition function} (\ref{HexagonPartition}) discussed in section \ref{TriangleSec} to generate spinning three point functions data and compare it with the perturbative results such as (\ref{C246}).

\subsection{Perturbative checks}
\label{perturbativechecks}
As derived in (\ref{recursionrelation}) the tree-level hexagon partition function can be written as a recursion relation. The minus signs introduced by gradding slightly change the expression (\ref{HtoZ}), but the physics behind it is still the same. One can fully determine the tree-level partition function by considering its decoupling poles and its behavior at infinity.

Both the recursion relation and its sum over partitions can be easily evaluated using for example \texttt{Mathematica}. Replacing the $\textbf{u}$, $\textbf{v}$ and $\textbf{w}$  by the Bethe roots for twist-2 operators with spins $J_1=2$, $J_2=4$ and $J_3=6$ we reproduce all the blue numbers in  (\ref{C246}).

One nice aspect of the integrability formalism is that, even though the amount of terms in the recursion relation and in the sum over partitions grows exponentially with the spins of the operators they do not grow with the twist of these operators. 

Therefore, the recursion relation above is a powerful tool to explore the corner of low spin and large twist three point functions. For example, the data for twist-10 operators with spins $J_1=2$, $J_2=4$ and $J_3=6$ can be easily evaluated. In particular for equal bridge lengths ($\ell_{ij}=5$) and Bethe roots
\begin{align*}
\mathbf{u} &= \{-0.0718891, 0.0718891 \}\,, \\
\mathbf{v} &= \{0.164766, 0.327921, 0.844103, 1.31232 \}\,, \\
\mathbf{w} &= \{0.296976, 0.491824, 0.756996, 1.13713, 2.62857, 3.91077\}\,,
\end{align*}
it reads

\begingroup\makeatletter\def\f@size{10}\check@mathfonts
\def\maketag@@@#1{\hbox{\m@th\large\normalfont#1}}%
\begin{align}
C(2,4,6) &=\langle 1  
     1\rangle  \langle 1  
     3\rangle  \langle 2  
     2\rangle^3  \langle 2  
     3\rangle  \langle 3  
     1\rangle  \langle 3  
     2\rangle  \langle 3  
     3\rangle^4 \left(\textcolor{blue}{-0.0000297465}\right)+\langle 1  
     2\rangle^2  \langle 2  
     1\rangle^2  \langle 2  
     2\rangle^2  \langle 3  
     3\rangle^6 \left(\textcolor{blue}{1.9812 \times 10^{-6}}\right)+ \nonumber \\ 
     &+ \langle 1  
     1\rangle  \langle 1  
     3\rangle  \langle 2  
     3\rangle^4  \langle 3  
     1\rangle  \langle 3  
     2\rangle^4  \langle 3  
     3\rangle  \left(\textcolor{blue}{0.00491993}\right)+\langle 1  
     3\rangle^2  \langle 2  
     2\rangle  \langle 2  
     3\rangle^3  \langle 3  
     1\rangle^2  \langle 3  
     2\rangle^3  \langle 3  
     3\rangle\left(\textcolor{blue}{-0.00216325}\right)+ \nonumber \\
     &+ \langle 1  
     2\rangle  \langle 1  
     3\rangle  \langle 2  
     1\rangle  \langle 2  
     3\rangle^3  \langle 3  
     1\rangle  \langle 3  
     2\rangle^3  \langle 3  
     3\rangle^2 \left(\textcolor{blue}{0.00999526}\right)+\langle 1  
     3\rangle^2  \langle 2  
     2\rangle^4  \langle 3  
     1\rangle^2  \langle 3  
     3\rangle^4\left(\textcolor{blue}{1.1861\times 10^{-6}}\right) + \nonumber \\
     &+\langle 1  
     2\rangle  \langle 1  
     3\rangle  \langle 2  
     1\rangle  \langle 2  
     2\rangle  \langle 2  
     3\rangle^2  \langle 3  
     1\rangle  \langle 3  
     2\rangle^2  \langle 3  
     3\rangle^3 \left(\textcolor{blue}{-0.00073364}\right) + \langle 1  
     1\rangle^2  \langle 2  
     2\rangle^4  \langle 3  
     3\rangle^6 \left(\textcolor{blue}{1.30467\times 10^{-7}}\right)+ \nonumber \\
     &+\langle 1  
     1\rangle  \langle 1  
     2\rangle  \langle 2  
     1\rangle  \langle 2  
     2\rangle^2  \langle 2  
     3\rangle  \langle 3  
     2\rangle  \langle 3  
     3\rangle^5 \left(\textcolor{blue}{-0.000071986}\right) + \langle 1  
     1\rangle^2  \langle 2  
     3\rangle^4  \langle 3  
     2\rangle^4  \langle 3  
     3\rangle^2 \left(\textcolor{blue}{-0.00144007}\right) + \nonumber \\
     &+\langle 1  
     3\rangle^2  \langle 2  
     2\rangle^3  \langle 2  
     3\rangle  \langle 3  
     1\rangle^2  \langle 3  
     2\rangle  \langle 3  
     3\rangle^3  \left(\textcolor{blue}{-0.000092995}\right)+\langle 1  
     1\rangle^2  \langle 2  
     2\rangle^2  \langle 2  
     3\rangle^2  \langle 3  
     2\rangle^2  \langle 3  
     3\rangle^4 \left(\textcolor{blue}{0.000522219}\right)+ \nonumber \\
     &+\langle 1  
     2\rangle^2  \langle 2  
     1\rangle^2  \langle 2  
     2\rangle  \langle 2  
     3\rangle  \langle 3  
     2\rangle  \langle 3  
     3\rangle^5  \left(\textcolor{blue}{0.0000727415}\right)+ \langle 1  
     1\rangle^2  \langle 2  
     2\rangle  \langle 2  
     3\rangle^3  \langle 3  
     2\rangle^3  \langle 3  
     3\rangle^3 \left(\textcolor{blue}{0.000908203}\right)+ \nonumber \\
     &+\langle 1  
     2\rangle  \langle 1  
     3\rangle  \langle 2  
     1\rangle  \langle 2  
     2\rangle^3  \langle 3  
     1\rangle  \langle 3  
     3\rangle^5 \left(\textcolor{blue}{-2.52289\times 10^{-6}}\right)+\langle 1  
     1\rangle  \langle 1  
     2\rangle  \langle 2  
     1\rangle  \langle 2  
     2\rangle^3  \langle 3  
     3\rangle^6 \left(\textcolor{blue}{-2.11167\times 10^{-6}}\right)+ \nonumber \\
     &+\langle 1  
     1\rangle  \langle 1  
     3\rangle  \langle 2  
     2\rangle^4  \langle 3  
     1\rangle  \langle 3  
     3\rangle^5  \left(\textcolor{blue}{-1.31656\times 10^{-6}}\right)+ 
  \langle 1  
     1\rangle  \langle 1  
     3\rangle  \langle 2  
     2\rangle  \langle 2  
     3\rangle^3  \langle 3  
     1\rangle  \langle 3  
     2\rangle^3  \langle 3  
     3\rangle^2  \left(\textcolor{blue}{-0.00342357}\right) + \nonumber \\
     &+\langle 1  
     3\rangle^2  \langle 2  
     3\rangle^4  \langle 3  
     1\rangle^2  \langle 3  
     2\rangle^4  \left(\textcolor{blue}{0.000683143}\right) + 
  \langle 1  
     1\rangle  \langle 1  
     2\rangle  \langle 2  
     1\rangle  \langle 2  
     2\rangle  \langle 2  
     3\rangle^2  \langle 3  
     2\rangle^2  \langle 3  
     3\rangle^4 \left(\textcolor{blue}{-0.00146526}\right) + \nonumber \\
     &+\langle 1  
     2\rangle  \langle 1  
     3\rangle  \langle 2  
     1\rangle  \langle 2  
     2\rangle^2  \langle 2  
     3\rangle  \langle 3  
     1\rangle  \langle 3  
     2\rangle  \langle 3  
     3\rangle^4  \left(\textcolor{blue}{0.000215642}\right) + 
  \langle 1  
     2\rangle^2  \langle 2  
     1\rangle^2  \langle 2  
     3\rangle^2  \langle 3  
     2\rangle^2  \langle 3  
     3\rangle^4 \left(\textcolor{blue}{0.00255232}\right) + \nonumber \\
     &+ \langle 1  
     3\rangle^2  \langle 2  
     2\rangle^2  \langle 2  
     3\rangle^2  \langle 3  
     1\rangle^2  \langle 3  
     2\rangle^2  \langle 3  
     3\rangle^2  \left(\textcolor{blue}{0.00212987}\right) + 
  \langle 1  
     1\rangle  \langle 1  
     2\rangle  \langle 2  
     1\rangle  \langle 2  
     3\rangle^3  \langle 3  
     2\rangle^3  \langle 3  
     3\rangle^3  \left(\textcolor{blue}{0.000540332}\right)+ \nonumber \\
     &+\langle 1  
     1\rangle  \langle 1  
     3\rangle  \langle 2  
     2\rangle^2  \langle 2  
     3\rangle^2  \langle 3  
     1\rangle  \langle 3  
     2\rangle^2  \langle 3  
     3\rangle^3  \left(\textcolor{blue}{-0.00103067}\right)+ 
  \langle 1  
     1\rangle^2  \langle 2  
     2\rangle^3  \langle 2  
     3\rangle  \langle 3  
     2\rangle  \langle 3  
     3\rangle^5  \left(\textcolor{blue}{9.51785\times 10^{-6}}\right)+ \nonumber \\
     &+\left(\langle 1  
     1\rangle ^2 \langle 2  
     3\rangle ^4 \langle 3  
     2\rangle ^4 \langle 3  
     3\rangle ^2-\langle 1  
      1\rangle  \langle 1  
     3\rangle  \langle 2  
     3\rangle ^4 \langle 3  
     1\rangle  \langle 3  
     2\rangle ^4 \langle 3  
     3\rangle\right) \left(\textcolor{blue}{0.0020167}\right)+ \nonumber \\
     & +\left( \langle 1  
     3\rangle ^2 \langle 2  
     2\rangle ^4 \langle 3  
     1\rangle ^2 \langle 3  
     3\rangle ^4-\langle 1  3\rangle ^2 \langle 2 
      2\rangle ^3 \langle 2  
     3\rangle  \langle 3  
     1\rangle ^2 \langle 3  
     2\rangle  \langle 3  
     3\rangle ^3\right)  \left(\textcolor{blue}{4.14448\times 10^{-6}}\right)+ \nonumber \\
     & + \left( \langle 1  
     2\rangle ^2 \langle 2  
     1\rangle ^2 \langle 2  
     2\rangle ^2 \langle 3  
     3\rangle ^6-\langle 1  2\rangle ^2 \langle 2 
      1\rangle ^2 \langle 2  
     2\rangle  \langle 2  
     3\rangle  \langle 3  
     2\rangle  \langle 3  
     3\rangle ^5\right) \left(\textcolor{blue}{6.5287 \times 10^{-6}}\right)+\dots\,,
   \label{C246Twisted}
\end{align}\endgroup

Beyond tree-level, the hexagon partition function is not a simple rational function of the rapidities. Lack of control over the singularities in other Riemann sheets prevent us from writing a recursion relation akin to (\ref{recursionrelation}) at loop order. We are forced to evaluate the scattering of the hexagon form factor via the explicit form of the hexagon partition function~(\ref{HexagonPartition}).

For general boundary conditions the one-loop evaluation is computationally demanding. One of the problems is that structure constants are given by linear combinations of multiple hexagons, which can be made explicit by combining (\ref{CinConfFrame}) and (\ref{CtoH}) to write
\begin{align}
\sum_{\ell_1,\ell_2,\ell_3} C^{J_1,J_2,J_3}_{\ell_1,\ell_2,\ell_3} \frac{\langle 1,1\rangle^{J_1-\ell_2-\ell_3}\langle 2,2\rangle^{J_2-\ell_1-\ell_3}\langle 3,3\rangle^{J_3-\ell_1-\ell_2}}{\langle 2, 3 \rangle^ {-\ell_1}\langle 3,2\rangle^{-\ell_1}\langle 1, 3 \rangle^{-\ell_2} \langle 3,1\rangle^{-\ell_2}\langle 1, 2 \rangle^{-\ell_3}\langle 2,1\rangle^{-\ell_3}} = \nonumber \\
= \mathcal{N}(\textbf{u})\mathcal{N}(\textbf{v})\mathcal{N}(\textbf{w})
\times \sum_{\substack{{a\, \cup \, \bar{a}=\textbf{u}}\\{b\, \cup \, \bar{b}=\textbf{v}}\\{c\, \cup \, \bar{c}=\textbf{w}}}} 
\omega_{\ell_{13}}(a,\bar{a})\omega_{\ell_{12}}(b,\bar{b})\omega_{\ell_{23}}(c,\bar{c})\times
\mathcal{H}(a,b,c) \times \mathcal{H}(\bar{a},{\bar{c}},{\bar{b}})\,.
\label{CtoHs}
\end{align}

When we fix the external polarizations $L_i$ and $R_i$ we must evaluate all the scaterrings, sum over partitions and in the end we obtain that a single glued hexagon is given by a combination of structure constants $C^{J_1,J_2,J_3}_{\ell_1,\ell_2,\ell_3}$. By considering several polarizations one can invert these relations obtaining single structure constants in terms of combinations of glued hexagons, see appendix \ref{inversion}.

For simplicity we will consider a particular polarization where the computation of a single hexagon gives us a simple combination of structure constants. This is the configuration where all the operators have the same polarizations (i.e all the excitations are chosen to be the longitudinal modes $\mathcal{D}^{1\dot{2}}$). In terms of spinors, these means $L_1=L_2=L_3$ and $R_1=R_2=R_3$, making the inner products of (\ref{CtoH}) equal. The tensor structure factors out and we get that
\begin{equation}
\langle 1,1 \rangle^{J_1+J_2+J_3} \sum_{\ell_1,\ell_2,\ell_3} C^{J_1,J_2,J_3}_{\ell_1,\ell_2,\ell_3}
\end{equation}
is given by a single glued hexagon where all the excitations have the same polarization. Note that, this is \textbf{not} an abelian configuration, once we cross the rapidities $\textbf{v}$ and $\textbf{w}$ to the top operator we end up with distinct excitations that do not scatter trivially.

For twist-2 operators we were able to evaluate this sum for arbitrary values of spins in a close formula
\begin{align}
\sum_{\ell_1,\ell_2,\ell_3} C^{J_1,J_2,J_3}_{\ell_1,\ell_2,\ell_3} & = \frac{\Gamma\left(1+J_1+J_2\right)\Gamma\left(1+J_1+J_3\right)\Gamma\left(1+J_2+J_3\right)}{\Gamma\left(1+J_1\right)\Gamma\left(1+J_2\right)\Gamma\left(1+J_3\right)\sqrt{\Gamma\left(1+2J_1\right)\Gamma\left(1+2J_2\right)\Gamma\left(1+2J_3\right)}}\times \nonumber \\
& \times \pFq{4}{3}{-J_1\text{,},-J_2\text{,},-J_3\text{,},-1-J_1-J_2-J_3}{-J_1-J_2\text{,},-J_1-J_3\text{,},-J_2-J_3}{1}
\label{treeSum}
\end{align}

By computing the equal polarizations hexagon for twist-2 and spin $J_1=2$, $J_2=4$ and $J_3=6$ we recovered this tree-level value from the hexagon partition function (\ref{HexagonPartition}) and also obtained the one-loop contribution
\begin{equation}
\sum_{\ell_1,\ell_2,\ell_3} C^{2,4,6}_{\ell_1,\ell_2,\ell_3} = \frac{75}{\sqrt{55}}-\frac{59077}{4620\sqrt{55}}g^2+\dots\,.
\end{equation}
which is reproduces the sum over all structure constants (all orange and cyan terms) appearing in the three point function (\ref{C246}).

\subsection{Abelian}
\label{abeliansection}
\begin{figure}[t]
\centering
\includegraphics[width=0.6\textwidth]{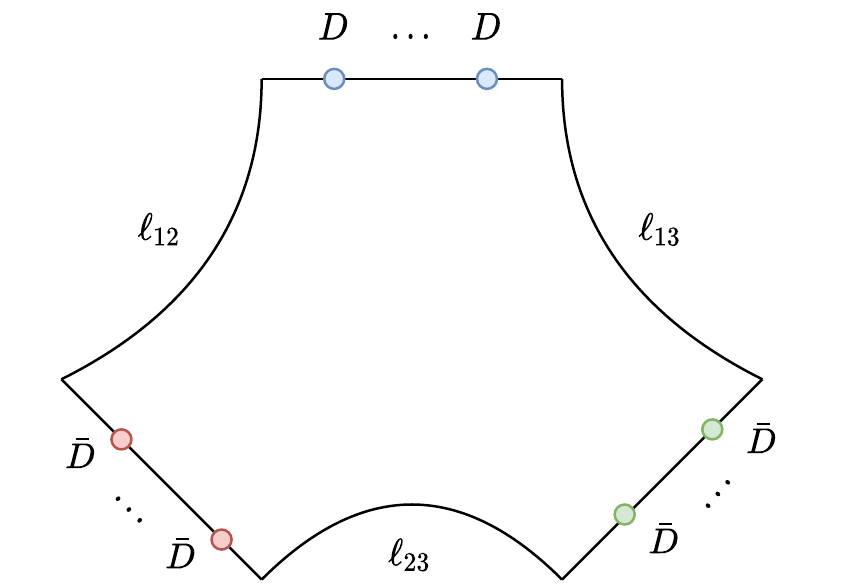}
\caption{The first operator is orthogonal to the other two operators, once we perform crossing we end up with identical excitations in the physical edge, which scatter trivially. This configuration corresponds to the abelian configuration $\mathbb{A}_1$. By considering the other configurations where the operators two and three are the ones orthogonal to the other operators we get the abelian structures $\mathbb{A}_2$ and $\mathbb{A}_3$, respectively.}
\label{figAbelian}
\end{figure}

As advertised in \ref{abelianpartition} there some polarizations which completely trivialize the scattering of the excitations. Since they are given by the abelian part of the hexagon partition function (\ref{HexagonPartition}), we denote them by \textbf{abelian} components. They are associated with three-point functions where one operator has orthogonal polarization to the other two, see figure (\ref{figAbelian}). By crossing these two operators we are left with three sets of excitations  with identical polarizations in the same physical edge, which in turn scatter trivially.

One of such configurations can be obtained by considering $R_1=L_2 =L_3$ and $L_1= R_2 = R_3$, which sets the inner products of (\ref{CinConfFrame}) to zero $\langle 1 2  \rangle = \langle 1 3 \rangle = 0$.  Moreover, $\langle 2 3 \rangle  \langle 3 2 \rangle = \langle 2 2 \rangle \langle 3 3 \rangle $, so the tensor structure factors out, resulting that the three point function
\begin{equation}
\langle 1,1 \rangle^{J_1}\langle 2,2 \rangle^{J_2}\langle 3,3 \rangle^{J_3} \sum_{\ell_1,\ell_2,\ell_3} C^{J_1,J_2,J_3}_{\ell_1,0,0}
\end{equation}
is given by an abelian hexagon. By permutation of the indices $1,2,3$ we have the other components
\begin{equation}
\mathbb{A}_1=\sum_{\ell_1} C^{J_1,J_2,J_3}_{\ell_1,0,0}\,, \quad \quad \mathbb{A}_2=\sum_{\ell_2} C^{J_1,J_2,J_3}_{0,\ell_2,0}\,, \quad \quad \mathbb{A}_3=\sum_{\ell_3}C^{J_1,J_2,J_3}_{0,0,\ell_3}
\end{equation}
which are structure constants given by hexagon form factors with purely abelian factors.

The scattering with these boundary conditions forces a trivial matrix part in (\ref{CtoH}) and yields the totally factorized contribution for the hexagon partition function shown in (\ref{abelian}) and reviewed here
\begin{align}
\mathcal{H}(u_a,v_b,w_c) & =  \langle 1, 1 \rangle^{a}  \langle 2, 2 \rangle^{b}   \langle 3, 3 \rangle^{c}  \times (-1)^{b + c}\prod_{v_i \in b}\prod_{u_j \in a}  \prod_{w_k \in c} h(v_i, u_j) h(u_j, w_k) h(v_i,w_k)\times \nonumber \\
& \times \prod_{\substack{{v_l \in b}\\{l<i}}}\prod_{\substack{{u_m \in a}\\{m<j}}}\prod_{\substack{{w_n \in c}\\{n<k}}} h(v_l,v_i)  h(u_m,u_j) h(w_n,w_k)\,,\label{HAbeliandef}
\end{align}
where the $h's$ above must be crossed according with their arguments, following (\ref{dynFactor}). 

These abelian form factors can be easily evaluated at any order in perturbation theory, in particular at one-loop and for twist-2 operators with spins $J_1=2$, $J_2=4$ and $J_3=6$ their sum over partitions reproduce the one-loop structure constants (colored in cyan) in expression (\ref{C246}).

It turns out that in the abelian case the sum over partitions can be performed analytically. The result (\ref{abelianfinal}) is a determinant formula for the three point function in the asymptotic regime valid for operators with arbitrary twist, which dependence enters only implicitly through the bridge lengths and the Bethe roots. This vastly reduces the computational effort required to evaluate the structure constants, and opens avenues to explore via integrability the large spin limit of the three point functions.

In this section we go over the result and derivation whose details and notation can be found in appendix \ref{PfaffianApendix}. The result for the case of one spinning operator was obtained in \cite{asymptoticfourpoint}. The key observation was that the sum over partitions of products of hexagon dynamical factors, 
$$H(u_{\bar{a}}, u_{\bar{a}}) \equiv \prod_{i<j \in \bar{a}} h(u_i,u_j) h(u_j,u_i),$$ could be interpreted as the expansion of a Fredholm pfaffian,
\beq
\sum_{\bar{a}\subset \mathbf{u}}(-1)^{|\bar{a}|} w'(u_{\bar{a}})H(u_{\bar{a}},u_{\bar{a}}) = pf(I-\mathcal{K})_{2J_1 \times 2J_1}. \label{pfaffianidentity}
\eeq
We define the matrix $\mathcal{K}$ in appendix \ref{PfaffianApendix}. Crucial for our purposes is that this holds for \textbf{any} factorized function of the rapidities $w'(u_{\bar{a}} ) = \prod_{i \in \bar{a}} w'(u_i)$.

 The asymptotic structure constant for three spinning operators (\ref{CtoHs}) can be rewritten in the abelian case as\footnote{For the rest of this section we will be explicit about the crossing kinematics $\circlearrowleft$, $\circlearrowright$.}
\begin{align}
\label{firststep} \mathbb{A}_1(J_1,J_2,J_3) = & \langle 1 1 \rangle^{J_1} \langle 2 2 \rangle^{J_2} \langle 3 3 \rangle^{J_3}  {\mathcal{N}(J_1)\mathcal{N}(J_2)\mathcal{N}(J_3)} h(\hat{u},\hat{u}) h(\hat{v},\hat{v}) h(\hat{w},\hat{w}) \\
&\times \sum_{\substack{{\color{red}{a \cup \bar{a}\subset \mathbf{u}}}\\ {\color{blue}{b\cup \bar{b} \subset \mathbf{v}}}\nonumber\\ {\color{blue}{c\cup \bar{c} \subset \mathbf{w} }}} }(-1)^{{\color{red}{|\bar{a}|}}{\color{blue}{+|\bar{b}|+|\bar{c}|}}} {\color{red}{e^u_{\bar{a}}}}{\color{blue}{e^v_{\bar{b}}e^w_{\bar{c}}}}{\color{red}{H(u_{\bar{a}},u_{\bar{a}})}} {\color{blue}{H(v_{\bar{b}},v_{\bar{b}}) H(w_{\bar{c}},w_{\bar{c}})  }}\\ & \times {\color{red}{h(v^{\circlearrowright}_{b},u_a) h(u_a,w^{\circlearrowleft}_{c})}}{\color{red}{h(w^{\circlearrowright}_{\bar{c}},u_{\bar{a}},) h(u_{\bar{a}},v^{\circlearrowleft}_{\bar{b}})}}  {\color{blue}{h(v^{\circlearrowright}_{b},w^{\circlearrowleft}_{c})  h(w^{\circlearrowright}_{\bar{c}} ,v^{\circlearrowleft}_{\bar{b}})}}, \nonumber
\end{align}
with $h(\hat{u},\hat{u}) \equiv \prod_{i<j} h(u_i,u_j)$, \beq e^u_{\bar{a}} \equiv \prod_{i \in \bar{a}} \frac{e^{i P(u_{i}) \ell_{13}}}{h(\hat{u},u_{i})},\qquad h(\hat{u},u_i) \equiv \prod_{j\neq i} h(u_j,u_i),  \qquad h(v^{\circlearrowright}_{b},u_a) \equiv \prod_{\substack{i\in a\\j\in b} }h(v^{\circlearrowright}_{j},u_i),  \nonumber\eeq and similar for the other terms. The red terms depend explicitly on both sets $a$, $\bar{a}$. However, by completing the factors $h(v^{\circlearrowright}_{b},u_a) h(u_a,w^{\circlearrowleft}_{c})$ to $h(v^{\circlearrowright}_{b},\hat{u}) h(\hat{u} ,w^{\circlearrowleft}_{c})$ and using the unitarity relation $h(x^\circlearrowright,y) h(y,x^\circlearrowleft) = 1$, it can be rewritten as
\begin{align}
{\color{red}{\text{red}}} &= \nonumber
{\color{blue}{h(v^{\circlearrowright}_{b},\hat{u}) h(\hat{u} ,w^{\circlearrowleft}_{c})}}\sum_{a \cup \bar{a} \subset \mathbf{u} }(-1)^{|\bar{a}|} e^u_{\bar{a}} H(u_{\bar{a}},u_{\bar{a}})    \frac{h(w^{\circlearrowright}_{\bar{c}},u_{\bar{a}}) h(u_{\bar{a}},v^{\circlearrowleft}_{\bar{b}})}{h(v^{\circlearrowright}_{b} ,u_{\bar{a}}) h(u_{\bar{a}},w^{\circlearrowleft}_{c}) } 
\\ &= \nonumber {\color{blue}{ h(v^{\circlearrowright}_{b},\hat{u}) h(\hat{u} ,w^{\circlearrowleft}_{c})}}\sum_{\bar{a} \subset \mathbf{u} }(-1)^{|\bar{a}|} \omega'(u_{\bar{a}}) H(u_{\bar{a}},u_{\bar{a}}) \nonumber,
\end{align}
with $\omega'(u_{\bar{a}}) = e^u_{\bar{a}}/\left(h(v^\circlearrowright,  u_{\bar{a}})h(u_{\bar{a}}, w^\circlearrowleft)\right)$, for which the pfaffian identity (\ref{pfaffianidentity}) applies.

Physically, from the point of view of $O_1$, the effect of the orthogonal operators $O_2$, $O_3$ is to create an effective background redefining the propagation factor $e^u_{\bar{a}}$ but leaving the interaction term $H(u_{\bar{a}},u_{\bar{a}})$ untouched. As we will see next, the same is true from the point of view of $O_2$, $O_3$ provided we interpret it as a larger operator with $J_2 + J_3$ excitations.

Plugging this result in (\ref{firststep}) leaves us with the blue terms,
\begin{align}
\label{blueterms}
{\color{blue}{\text{blue}}} = \sum_{\substack{ b\cup \bar{b} \subset \mathbf{v}\\ c\cup \bar{c} \subset \mathbf{w} } }(-1)^{|\bar{b}|+|\bar{c}|} e^v_{\bar{b}}e^w_{\bar{c}}H(v_{\bar{b}},v_{\bar{b}}) H(w_{\bar{c}},w_{\bar{c}})  h(v^{\circlearrowright}_{b},\hat{u}) h(\hat{u} ,w^{\circlearrowleft}_{c})h(v^{\circlearrowright}_{b},w^{\circlearrowleft}_{c}) h(w^{\circlearrowright}_{\bar{c}} ,v^{\circlearrowleft}_{\bar{b}}). \end{align}
Once again completing to the overall factors and using the unitarity identity, we can reduce\footnote{See appendix (\ref{PfaffianApendix}) for details.} the summand to functions of $\bar{b}$, $\bar{c}$ only,
\beq  {\color{blue}{\text{blue}}} = h(\hat{v}^{\circlearrowright},\hat{w}^{\circlearrowleft})  h(\hat{v}^{\circlearrowright},\hat{u}) h(\hat{u} ,\hat{w}^{\circlearrowleft} )\sum_{\substack{ b\cup \bar{b} \subset \mathbf{v}\\ c\cup \bar{c} \subset \mathbf{w} } }(-1)^{|\bar{b}|+|\bar{c}|} \omega'(v_{\bar{b}}^{\circlearrowright})\omega'(w_{\bar{c}}) H(v_{\bar{b}}^{\circlearrowright},v_{\bar{b}}^{\circlearrowright}) H(w_{\bar{c}},w_{\bar{c}})   H(v^{\circlearrowright}_{\bar{b}},w_{\bar{c}})   \label{blueresult}, \eeq
with $\omega'(v_{\bar{b}}^{\circlearrowright})$, $\omega'(w_{\bar{c}})$ defined in appendix (\ref{PfaffianApendix}). To conclude, one must reinterpret the double sum over partition as a single sum over the partition of the set $\mathbf{z} = \mathbf{v}^\circlearrowright \cup \mathbf{w}$. One can then write  
\beq
 {\color{blue}{\text{blue}}}  = h(\hat{v}^{\circlearrowright},\hat{w}^{\circlearrowleft})  h(\hat{v}^{\circlearrowright},\hat{u}) h(\hat{u} ,\hat{w}^{\circlearrowleft} )\sum_{d\cup \bar{d} \subset \mathbf{z} }(-1)^{|\bar{d}|} \omega'(z_{\bar{d}}) H(z_{\bar{d}},z_{\bar{d}})   \nonumber
\eeq
which can be cast as a pfaffian through (\ref{pfaffianidentity}). 
  
  The final result is
  \begin{align}
\mathbb{A}_1(J_1,J_2,J_3) & = \langle 1 1 \rangle^{J_1} \langle 2 2 \rangle^{J_2} \langle 3 3 \rangle^{J_3}  h(\hat{v}^{\circlearrowright},\hat{u})  h(\hat{u},\hat{w}^{\circlearrowleft})h(\hat{v}^{\circlearrowright},\hat{w}^{\circlearrowleft}) h(\hat{v},\hat{v})h(\hat{w},\hat{w})h(\hat{u},\hat{u})  \times \nonumber \\
 &  {\mathcal{N}(J_1)\mathcal{N}(J_2)\mathcal{N}(J_3)} \times pf\left(\textit{I} - \mathcal{K}\right)_{2J_1 \times 2J_1 } pf\left(\textit{I} - \mathcal{K}'\right)_{2(J_2+J_3) \times 2(J_2+J_3) },  \label{abelianfinal}
 \end{align}
 with the matrices $\mathcal{K}$ and $\mathcal{K}'$ defined in appendix (\ref{PfaffianApendix}).

As discussed above, these formulas express the abelian three point functions at finite coupling in the asymptotic regime, that is, ignoring finite volume corrections coming from excitations wrapping around the seams of the pair of pants. These mirror corrections can be properly ignored at weak coupling when we consider three point functions of large twist operators so that all bridge numbers $\ell_{ij}$ are large. Due to the Boltzmann suppression of propagation over large distances $\ell_{ij}$, mirror particles only start to contribute at $(\ell_{ij} +1)$ loops.

 For example, when we consider operators of classical twist $\tau_1 = \tau_2 = \tau_3 = 6$ and bridge lengths $\ell_{ij} = 3$, equation (\ref{abelianfinal}) provides exact formulas up to three loops. As an example, for spins $J_1 = 6$, $J_2 =4$, $J_3 = 2$ and bethe roots (at tree-level) 
\begin{align*}
\mathbf{u} &= \{-3.34538, -2.2266, -1.45396, -0.894775, -0.500817, -0.22559\}\,, \\
\mathbf{v} &= \{-2.18949, -1.24366, -0.66142, 0.163878 \}\,, \\
\mathbf{w} &= \{-1.03826, 1.03826 \}\,,
\end{align*}
we obtain\footnote{We keep the $\zeta(3)$ from the dressing phase explicit.},
\begin{align*}
C_\text{abelian}(6,4,2) =  &0.000087168875373411949253081221196521450200295689674319409 \\
-&0.0010941423754946036043705421214308022132833891655992081\, g^2 \\
+ &0.0024639150636611583478721259566761789123679977153044975 \,g^4  \\
 + &(0.15149466528731340172164338220385987881409857001459463    \\
+ &0.0099644970255952430277488412973707304997053789074945953 \times \zeta(3)) g^6 .
\end{align*}
We hope that the novel data generated through this result will be useful in future integrability explorations. 

\subsection{A two loop check}
\label{CtoH2loops}

An important question is if the asymptotic part is sufficient to capture the leading contribution, in the large spin limit, of three point functions with more than one non-BPS operator. The goal of this section is to lay out the main ingredients that are necessary to compute the first correction to the asymptotic part of the OPE coefficient in the Abelian polarization. Most of the discussion is valid for operators with any spin but we compute explicitly only the finite size correction of the OPE coefficient of two Konishi operators and one BPS leaving a more general analysis(including the large spin limit mentioned above) to the future.  
These new effects are obtained by gluing the two hexagons along the three seams (from now on referred to as mirror edges) to form a pair of pants. The gluing procedure is achieved by dressing the mirror edges with  multiparticle states and integrating over their rapidities.  The multiparticle states are labeled by boundstate number $n$ as explained \cite{hexagons,Basso:2015eqa,Eden:2015ija,Basso:2017muf}. 

\begin{figure}[t]
\centering
\includegraphics[scale=0.20]{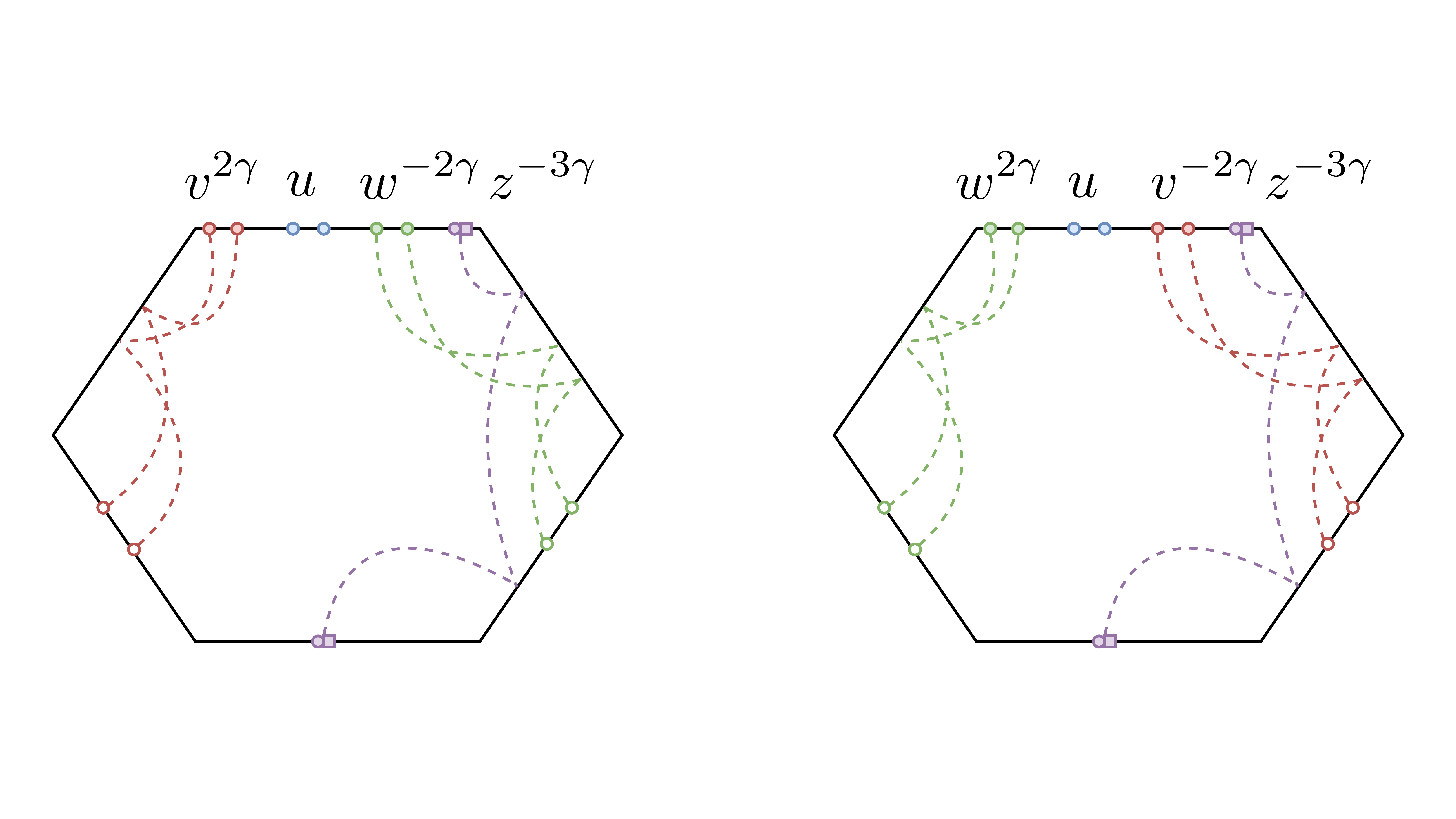}
\caption{Single mirror particle contribution to a three point function with three spinning operators.  The mirror particle would correspond to a bottom contribution (in the notation of \cite{hexagons}) if the operators $2 ,3$ were BPS and adjacent if the the $1$ and $2$ or $3$ were absent.  }
\label{FiniteSize1}
\end{figure}

One crucial but important detail is that the bound state particles live in the mirror edges. This might seem innocuous but is the reason why their contribution only starts at two loops. To have a better grasp on this matter let us look at the integrand of just one boundstate particle in the hexagon formalism 
\begin{align}
\textrm{int}&=\sum_{\substack{{a\, \cup \, \bar{a}=\textbf{u}}\\{b\, \cup \, \bar{b}=\textbf{v}}\\{c\, \cup \, \bar{c}=\textbf{w}}\\X}}   \mu (z^{\gamma}) e^{ip_{z}^{\gamma}} \omega_{l_{13}}(a,\bar{a})\omega_{l_{23}}(c,\bar{c})\omega_{l_{12}}(b,\bar{b}) \mathcal{H}(b^{2\gamma},a,c^{-2\gamma},z^{-3\gamma})\mathcal{H}(\bar{c}^{2\gamma},\bar{a},\bar{b}^{-2\gamma},z^{-3\gamma})\label{eq:integrandfinitesizecorr1}
\end{align}
where $X$ is a particle in the mirror theory with rapidity $z$ and 
\begin{align}
\mu(z^{\gamma})= \frac{n(x^{[+n]}x^{[-n]})^2}{g^2(x^{[+n]}x^{[-n]}-1)^2((x^{[+n]})^2-1)(1-(x^{[-n]})^2)},  \  \ \ e^{ip_{u^\gamma}}=\frac{1}{x^{[n]}x^{[-n]}}
,\end{align}
with $n$ being the boundstate number. Since the boundstate particle lives in the mirror edge we need to do and off number of $\gamma$  rotation. In this case the expressions for $\mu(z^{\gamma})$ and $e^{ip_{u^\gamma}}$  start with $g^2$ at weak coupling. Physically, this is just saying that we pay a coupling for the mirror particle to travel from the front hexagon to the back. For this reason the finite size correction for twist two operators only makes an appearance at the two loop level.  Fortunately there is perturbative data computed using traditional Feynman integrals that we can compare against \cite{Bianchi:2021wre}. The rest of this section is devoted to obtain the finite size correction at two loops and for two spin two operators and match it with the result obtained in \cite{Bianchi:2021wre}.

The integrand (\ref{eq:integrandfinitesizecorr1}) is very similar to the hexagon partition function, as it is the scattering of three sets of particles in a pairwise fashion, but now with an additional bound state particle arising from the mirror contributions. 

\begin{figure}[t!]
\centering
\includegraphics[scale=0.15]{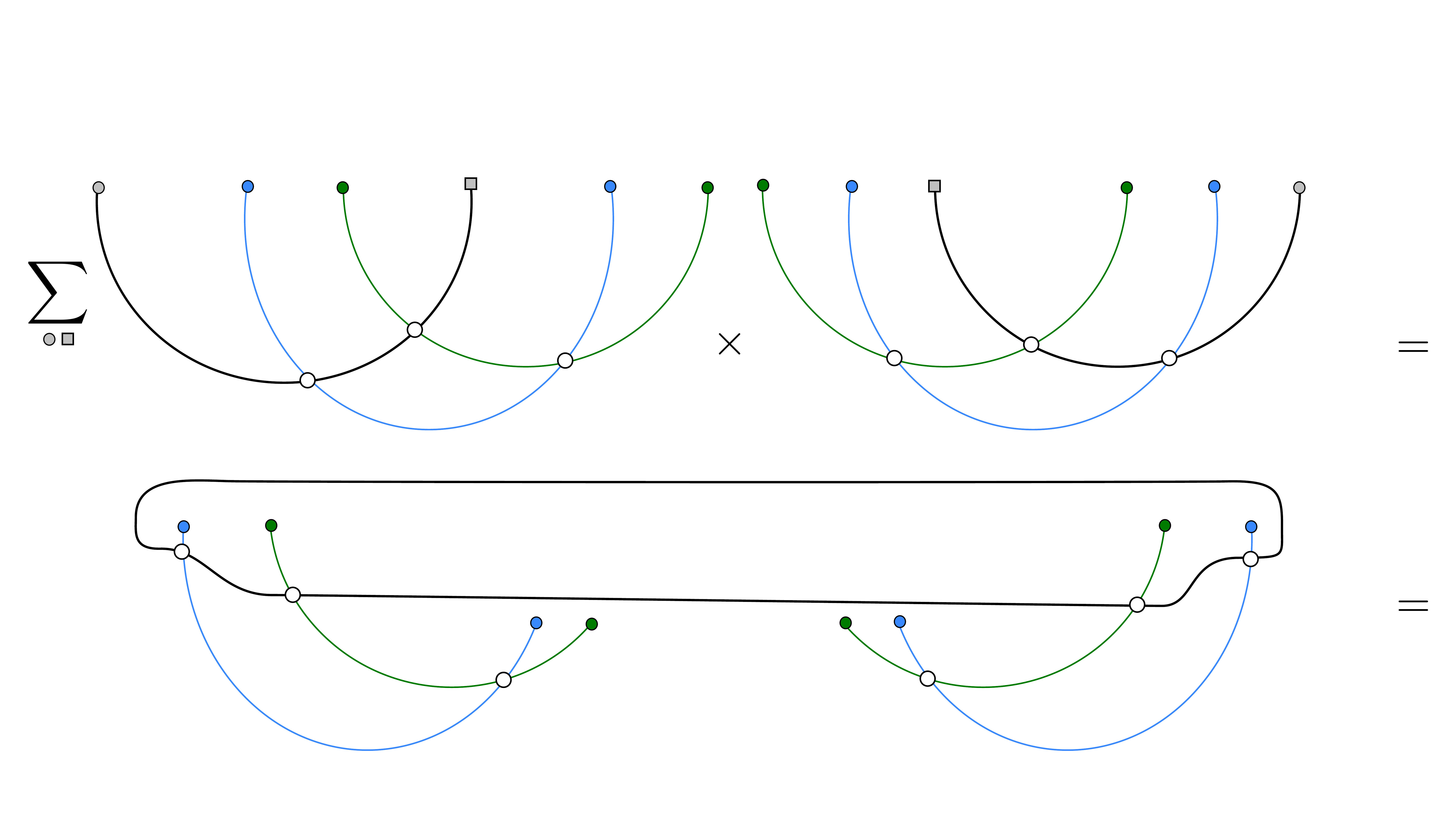}
\caption{ The sum over the flavour indices of the mirror particle states in the matrix part of the hexagon form factor is proportional to a transfer matrix. This is precisely the same mechanism present in the OPE with just one spinning operator. }
\label{fig:TransferMatrixPic}
\end{figure}

There are several simplifications for the Abelian polarization,  for example the scattering between the three sets of particles $\textbf{u}$, $\textbf{v}$ and $\textbf{w}$ is simple. If it was not for the boundstate particles, the whole matrix part would be trivial. However, as depicted in figure \ref{fig:TransferMatrixPic}, in this abelian configuration the boundstate scattering can be recasted as a transfer matrix \cite{hexagons},
\begin{equation}
T_{a}(z) = \sum_{n=-1}^{1}(3n^2-2)\prod_{m=0}^{n}\frac{R^{(+)}(z^{[2m-a]})}{R^{(-)}(z^{[2m-a]})} \sum_{j=\frac{2-a}{2}}^{\frac{a-2n}{2}}\prod_{k=j+n}^{\frac{a-2}{2}}\frac{R^{(+)}(z^{[2n-2k]})B^{(+)}(z^{[-2k]})}{R^{(-)}(z^{[2n-2k]})B^{(-)}(z^{[-2k]})}\label{eq:transfermatrixdefinition}\, ,
\end{equation}
where
\begin{equation}
R^{(\pm)}(z) = \prod_{j}(x(z)-x^{\mp}_{j})\, , \,\,\, B^{(\pm)}(z) = \prod_{j}\left(\frac{1}{x(z)}-x^{\mp}_{j}\right)\, .
\end{equation}
and the product in $j$ should be taken over all physical particles.  Notice that some of these  particles are mirror rotated as can be seen in figure \ref{FiniteSize1} and for this reason the transfer matrix that shows up in the integrand (\ref{eq:integrandfinitesizecorr1}) is not exactly (\ref{eq:transfermatrixdefinition}). Instead, the integrand will be written in term of  $T_{n}^{\circlearrowright\circlearrowleft}(z)$ which is defined by 
\begin{align}
T_{n}^{\circlearrowright\circlearrowleft}(z)  = T_n(z)\Big|_{x^{[-a]}\to \frac{1}{x^{[-a]}} ,\, x^{\pm}_j \to \frac{1}{x^{\pm}_j}} \ \ \  j\in {\bf{v} \cup \bf{w}}. 
\end{align}

The integrand (\ref{eq:integrandfinitesizecorr1}) can then be written in a compact form as 
\begin{align}
&\textrm{int}(z)=\frac{T_{n}^{\circlearrowright\circlearrowleft}(z)\mu (z^{\gamma}) e^{ip_{z}^{\gamma}} }{h_n(z^{-\gamma},{\bf{v}})h_n(z^{\gamma},{\bf{u}}) }h_n({\bf{w}},z^{-\gamma}) \sum_{\substack{{a\, \cup \, \bar{a}=\textbf{u}}\\{b\, \cup \, \bar{b}=\textbf{v}}\\{c\, \cup \, \bar{c}=\textbf{w}}}}   \omega_{l_{23}}'(c,\bar{c})\omega_{l_{13}}(a,\bar{a})\omega_{l_{12}}'(b,\bar{b})\times \nonumber\\
&\times h(b^{2\gamma},a)h(b^{4\gamma},c)h(a,c^{-2\gamma})h(\bar{c}^{2\gamma},\bar{a})h(\bar{c}^{4\gamma},\bar{b}_2)h(\bar{a},\bar{b}^{-2\gamma})\prod_{i=1}^3h(w_{i},w_{i})h(\bar{w}_{i},\bar{w}_{i}),\label{eq:integrandfinitesize}
\end{align}
with $\omega'$, $p_n$ and $h_{n}$ defined by
\begin{align}
\omega^\prime_{\ell_{23}}(c,\bar{c}) = \omega_{\ell_{23}}(c,c)p_n\left(\bar{c}^{4\gamma},z^{-\gamma}\right)\,, \quad \omega^\prime_{\ell_{12}}(b,\bar{b}) = \omega_{\ell_{12}}(b,b)p_n\left(\bar{b},z^{-\gamma}\right)\,,& \\ 
p_n(u,v) = h_n(u,v)h_n(v,u) =  \frac{(u-v)^2+\frac{(n-1)^2}{4}}{(u-v)^2+\frac{(n+1)^2}{4}}\left(\frac{1-\frac{1}{y^{-}x^{[+n]}}}{1-\frac{1}{y^{-}x^{[-n]}}}\frac{1-\frac{1}{y^{+}x^{[-n]}}}{1-\frac{1}{y^{+}x^{[+n]}}}\right)^2\,,& \\
S_n(u,v) = \frac{1}{\sigma_n^2(u,v)}\frac{(u-v+i\frac{n-1}{2})(u-v+i\frac{n+1}{2})}{(u-v-i\frac{n-1}{2})(u-v-i\frac{n+1}{2})}\prod_{k=-\frac{n-1}{2}}^{\frac{n-1}{2}} \left(\frac{1-\frac{1}{y^+x^{[-2k-1]}}}{1-\frac{1}{y^-x^{[+2k+1]}}}\right)^2&
\end{align}
and $\sigma_{n}(u,v)$ is the (fused) BES dressing phase \cite{BES}. 
Let us remark that these formulas should be valid for operators with any spin but only in the Abelian polarization.

\begin{figure}[t!]
\centering
\includegraphics[width=1\textwidth]{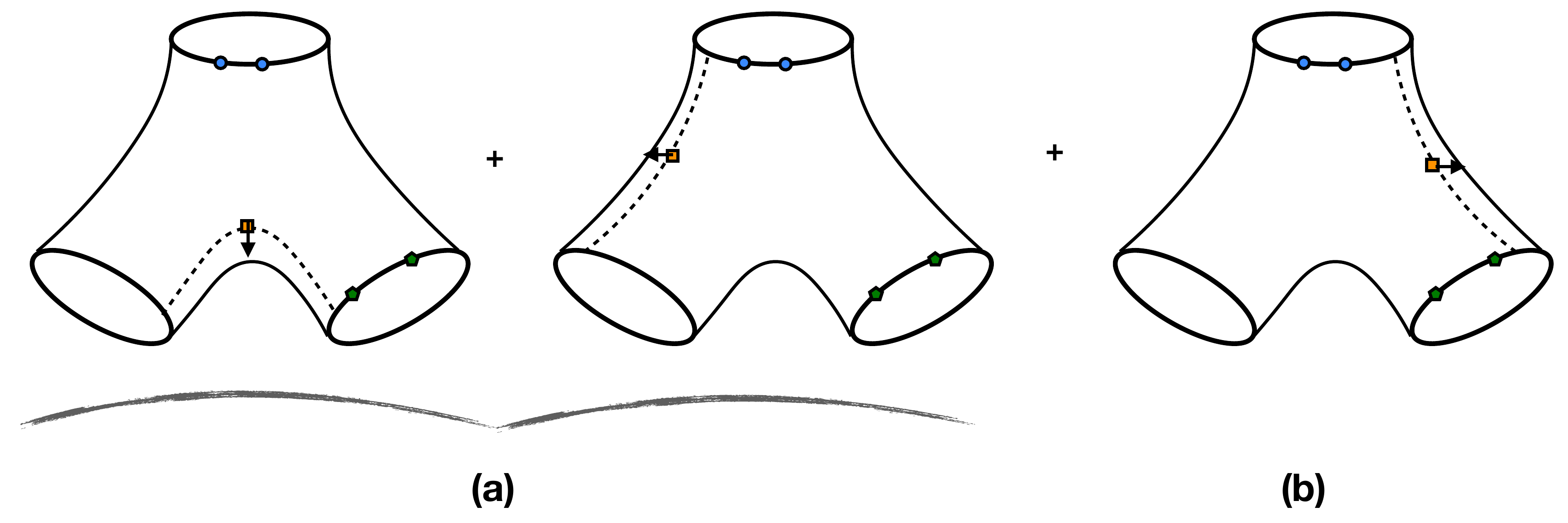}
\caption{For two non-BPS operators there is one mirror edge that is distinct from the other two. It turns out that up to two loops only the finite size corrections (a) contribute while the finite size correction (b) will only starts at three loop order.   }
\label{AdjacentBotttom}
\end{figure} 

For simplicity we focus on three point functions with two spinning operators $J_1=J_2=2$ and one BPS and compare it with perturbative results obtained in \cite{Bianchi:2021wre}. Obviously, this choice breaks the symmetry between the mirror edges since one of them will be between the two non-BPS operators, as can be seen in figure \ref{AdjacentBotttom}. It is straightforward to check by expanding at weak coupling that we only need to take into consideration the finite size corrections represented in \ref{AdjacentBotttom}.a. 



The building blocks to get this contribution are very explicit and now it is just a matter of evaluating them at weak coupling after taking the appropriate monodromies. After doing this we obtain
\begin{align}
\textrm{int}(z) = {\scriptstyle{\frac{55296 (-1)^{2/3} n^2 g^4 \left(12 z^2-n^2\right) \left(3 n^2+12 z^2-4\right)}{\left(n^2+4 z^2\right)^3 \left(216 n^6 \left(6 z^2-1\right)+432 n^4 \left(18 z^4-2 z^2+1\right)+384 n^2 \left(54 z^6+9 z^4-15 z^2-1\right)+81 n^8+256 \left(9 z^4+3 z^2+1\right)^2\right)}}}\label{eq:expandedintegrandweakcoupling}
\end{align}
where we used the Konishi Bethe roots for the two spinning operators 
\begin{equation}
\textbf{u}=\textbf{v}=\left(-\frac{1}{2\sqrt{3}},\frac{1}{2\sqrt{3}}\right)\,.
\end{equation} 
The OPE coefficient\footnote{One might be worried about the $(-1)^\frac{2}{3}$ in  (\ref{eq:expandedintegrandweakcoupling})  because it can give rise to an imaginary part in the structure constant. However this does not pose an issue since this factor is removed with the normalization factor $\mathcal{N}(2) ^2=  \frac{(-1)^{-\frac{2}{3}}}{54}$.} for two spin two operators is given by at two loops
\begin{equation}
 \sum_{\ell=0}^2C_{2,2,0}^{\ell} = \mathcal{N}(2)\mathcal{N}(2) \left(\textrm{{\bf{asymptotic} }}_{\textrm{{\bf{Abelian} }}} + 2\sum_{n=1}^\infty \int dz \textrm{int}(z) \right).
\end{equation}
where \textbf{asymptotic} are the abelian structures given by the hexagon partition function (\ref{HAbeliandef}) or by the pfaffian (\ref{abelianfinal}) and the factor of two in front of the integral takes into account the two possibilities to add mirror particles as shown in figure \ref{AdjacentBotttom}.a. 

The integral over the rapidity $z$, that can be done by picking residues and the sum over the boundstate $n$ can be evaluated without much effort.  After adding everything up we obtain
\begin{align}
 \sum_{\ell=0}^2C_{2,2,0}^{\ell} = \frac{1}{6}\big(1-12 g^2 +147 g^4\big)
\end{align}
which matches exactly the result obtained in \cite{Bianchi:2021wre}. 

It should be possible to obtain the finite size correction for other spins (in the Abelian polarization). The only issue that might occur is that the integral/sum needs to be regularized. However, we expect that the HPL method of \cite{Basso:2015eqa} can be used to regularize efficiently the sums in a similar fashion as in one spinning case. Along these lines it would be interesting to extend the results of \cite{Basso:2022nny} to our setting. 

Obtaining finite size corrections for other polarizations is somewhat harder but should be doable at least at leading order in the coupling and for small spins. The main issue is that the sum over mirror particles in (\ref{eq:integrandfinitesizecorr1}) is more involved. It would be interesting to study this further in the future.

\section{Discussion}
This paper reduces the computation of asymptotic three point functions of three spinning operators in $\mathcal{N}=4$ SYM to the statistical mechanical problem of computing the partition function of a system of the Hubbard type on a Kagome-like lattice whose boundary conditions are determined by the polarizations and quantum numbers of the spinning operators, equations (\ref{HexagonPartition}, \ref{CtoH}, \ref{HtoZ}). The analytic structure of this partition function is inherited from the vertex, Beisert's SU(2|2) extended S-matrix \cite{Beisert1}, and therefore is extremely rich. Its singularities are determined by the spectrum of the dual world-sheet theory and have clear physical interpretations: particles decoupling, annihilating, fusing into bound states, and others.

In the limit of weak coupling the vertex reduces to the rational type and the partition function can be solved. We do so in this work by exploiting the analyticity of the partition function. After stripping out overall factors, the partition function in the rational limit is a meromorphic function of the rapidities with no bound-state poles. It can be then solved recursively by concatenating the various decoupling poles in each of the rapidities. This is the result (\ref{recursionrelation}). From the point of view of the gauge theory, it provides an efficient way of generating tree-level structure constant data for spinning operators of arbitrary twist. 

It would be fascinating to solve this partition function away from weak coupling. From the point of view of analyticity, we now have functions on a multi-sheeted Riemann surface with multiple additional fusion poles whose residues should be related to smaller partition functions involving bound-state lines. From the statistical mechanics viewpoint, similar integrable systems were solved in a more traditional fashion. For example, in \cite{baxter} Baxter solves the thermodynamic partition function of a 8-vertex model defined on the faces of a Kagome lattice by matching it to the computation in the standard square lattice. His construction is the same behind the integrability of the fishnet model by Zamolodchikov \cite{Zpaper}. Exploring solid-on-solids dualities of this type might lead to a direct solution in terms of the transfer-matrices and functional equations. 

Once the statistical mechanics problem is solved, one must still resolve the sum over partitions (\ref{CtoH}), the complexity of which grows rapidly at large \textit{but finite} spin\footnote{In the strict large spin limit we hope to have alternative techniques, see discussion below.}. Turns out this problem can be solved, at any value of the coupling, for certain boundary conditions of the partition function. When two operators are parallel and orthogonal to the third the hexagon becomes a simple abelian factor, the sum over partitions of which can be cast in pfaffian form. This is the result (\ref{abelianfinal}). It provides an efficient way of generating some high-loop data for structure constants of high twist spinning operators.\footnote{Some corners of the cross ratios space parametrized by the null snowflake OPE limit of the 6-pt function should be controlled by the abelian structure constants. Having complete control over those, we could therefore generalize \cite{asymptoticfourpoint} to the higher point case and provide valuable boundary data for an eventual bootstrap approach to fix these correlators \cite{enrico, frank, vasco}.
} Note: for each generic three spinning operators, the pfaffian formulas provide three independent structure constants, one for each choice of orthogonal operator. The three point function must therefore interpolate between these three highly non-trivial pfaffian formulas involving matrices of different dimensionality as the polarization vectors vary. It would be great to identify non-trivial objects linear in the tensors structures (\ref{CinConfFrame}) that performs such interpolation, as they might be relevant for describing the full three point function of spinning operators in $\mathcal{N} = 4$ SYM.

\begin{figure}[t!]
\centering
\includegraphics[width=0.75\textwidth]{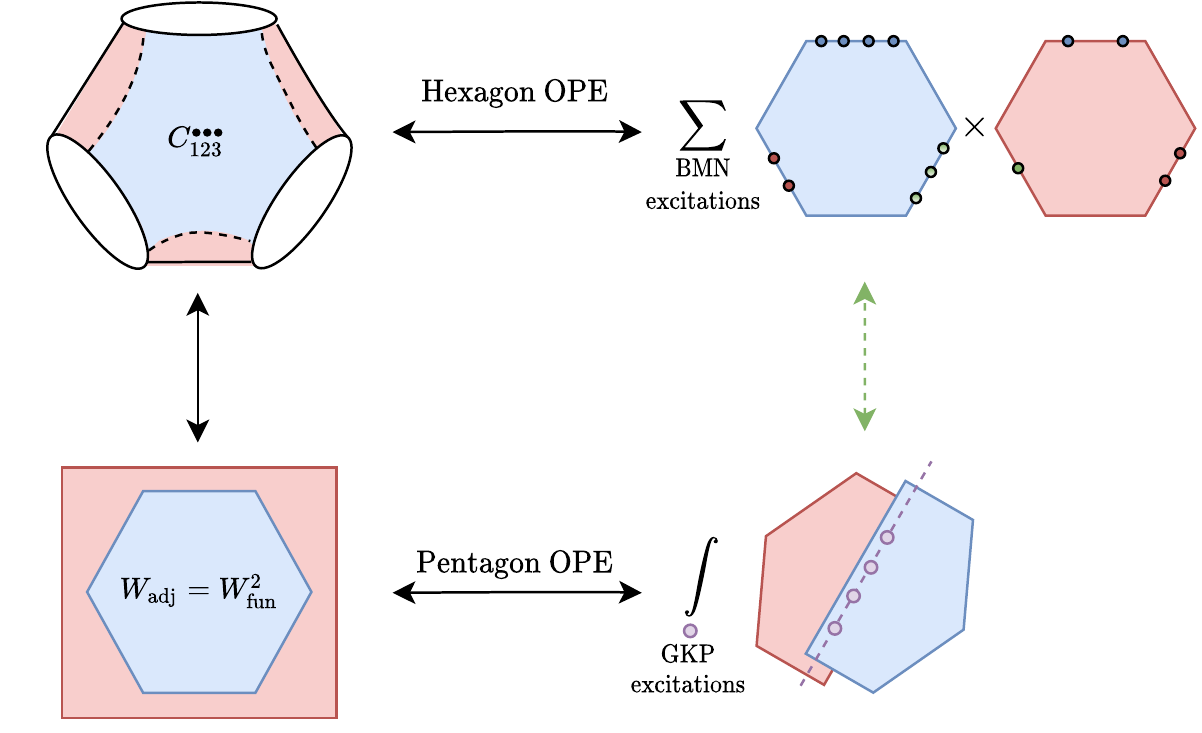}
\caption{\textbf{Top arrow:} this work, decomposition of the spinning three point functions into polarized hexagons. \textbf{Left arrow:} large spin three-point functions duality with null Wilson loops \cite{Multi}. \textbf{Bottom arrow:} decomposition of null Wilson loops into integrable pentagons \cite{POPE}. \textbf{Right arrow:} future work, how the integrable hexagons and integrable pentagons are related in the large spin limit.}
\label{figDiscussion}
\end{figure}

At large spin the asymptotic structure constant should provide the exact three point function of the gauge theory, finite size corrections being power-law suppressed in spin. In \cite{paper1} we considered the null-hexagon limit of six point functions of $\mathbf{20}'$ operators and its respective snowflake OPE decomposition in terms of three large spin operators to derive a map between the expectation value of null hexagonal Wilson loops (WL) and the large spin limit of the three point functions considered in this paper. Combining \cite{paper1} and this work, we can write a sharp formula expressing the expectation value of the null WL in any geometrical configuration to the hexagons expansion:
\beq
 \mathbb{W}(U_1, U_2, U_3) = \lim_{J_i, \ell_i \rightarrow \infty}  \frac{\mathcal{N_\lambda} \mathcal{N}_\textbf{u} \mathcal{N}_\textbf{v}\mathcal{N}_\textbf{w} }{\left(C^{\ell_1,\ell_2,\ell_3}_{J_1,J_2,J_3}\right)_
{\text{tree}}}
\times 
\prod_{i=1}^3 \left(\frac{J_i \ell_i}{2\ell_{i+1}\ell_{i-1}}\right)^{-\frac{\gamma_i}{2}}
\times
\sum_{\substack{{a\, \cup \, \bar{a}=\textbf{u}}\\ {b\, \cup \, \bar{b}=\textbf{v}}\\ {c\, \cup \, \bar{c}=\textbf{w}}}} 
\omega_{\ell_{13}}\omega_{\ell_{12}}\omega_{\ell_{23}} \times
\mathcal{H}_{a,b,c}  \mathcal{H}_{\bar{a},{\bar{c}},{\bar{b}}} \label{mapintegrability}
\eeq
where the ratios of $\ell_i$ and $J_i$ are kept fixed and determined in terms of the null hexagon cross ratios  $U_i$ through (10) of \cite{paper1}; the boundary conditions of the hexagon partition function~$\mathcal{H}$ are fixed in terms of $U_i$ through (37) of \cite{paper1}. Above, $\gamma_i$ are the anomalous dimensions of operator $O_i$, while $\mathcal{N}_\lambda$ is a spin independent normalization constant to be determined.

Null WL can be decomposed as an expansion around the collinear limit by the Pentagon OPE, higher energy excitations of the GKP vacuum controlling the geometric expansion \cite{POPE}. Equation (\ref{mapintegrability}) can therefore be understood as a map between hexagons and pentagons, see figure \ref{figDiscussion}. Since the adjoint Wilson loop $\mathbb{W}$ is the square of the fundamental Wilson loop in the large $N$ limit and since the latter is obtained gluing two pentagons $\mathcal{P}$ together we can cast the sharp equation (\ref{mapintegrability}) as the simplified slogan
\beq
\(\sum \mathcal{P} \times \mathcal{P} \)^2 = \lim_{\texttt{large spin}} \sum \mathcal{H} \times \mathcal{H}  \,.
\eeq
Explicitly uncovering this relation is a problem for the future which could lead to a more universal integrability framework for $\mathcal{N}=4$ SYM. 
More generally, one should ask if (\ref{mapintegrability}) could lead to an alternative effective way of computing WL physics and their dual gluon scattering amplitudes \cite{d1,lance}. The answer to this question depends on understanding how to simplify the large spin limit of the computations discussed in this work. The thermodynamic limit of the hexagon partition function is therefore of special importance. 
The matrix-models inspired methods of \cite{clustering,tunel,classicalkostov} should also be useful. 

It is also possible that instead of simplifying the building blocks arising in the summand in (\ref{mapintegrability}) we should discard all these partition functions and look for new tricks to compute directly the full sum over partitions. That would be bitter sweet.\footnote{Even if that turns out to be the case, these (sums over) hexagon partition functions will always be useful to produce important data to test such scenario specially in perturbation theory.}

 \section*{Acknowledgements}
We would like to thank Benjamin Basso, Vladimir Kazakov, Jo\~{a}o Caetano, Gabriel Lefundes, Matheus Fabri, Tiago Fleury, Shota Komatsu for illuminating discussions. Research at the Perimeter Institute is supported in part by the Government of Canada 
through NSERC and by the Province of Ontario through MRI. We also thank  Porto University for the organization of the 2022 Bootstrap conference in Porto, during which this project was completed. This work was additionally 
supported by a grant from the Simons Foundation (Simons Collaboration on the Nonperturbative Bootstrap \#488661) and ICTP-SAIFR FAPESP grant 2016/01343-7 and FAPESP grant 2017/03303-1. The work of P.V. was partially supported by the Sao Paulo Research Foundation (FAPESP) under Grant No. 2019/24277-8. The work of C.B. was supported by the Sao Paulo Research Foundation (FAPESP) under Grant No. 2018/25180-5. V.G. is supported by Simons Foundation grants \#488637 (Simons collaboration
on the non-perturbative bootstrap). Centro de F\'{i}sica do Porto is partially funded by Funda\c{c}\~{a}o para a Ci\^{e}ncia e Tecnologia (FCT) under the grant UID04650-FCUP. This research was supported in part by the National Science Foundation under Grant No. NSF PHY-1748958.

\appendix 

\section{Spinors}
\label{spinors}

In section \ref{SecCtoH} we parametrize spinning operators through their left- and right-handed polarization spinors ${L_i}_\alpha $ and ${R_i}_{\dot{\alpha}}$. These are related to the polarization vectors by
\begin{equation}
\epsilon_i^\mu = {R_i}_{\dot{\beta}}  \left(\bar{\sigma}^\mu\right)^{\dot{\beta} \alpha} {L_i}_\alpha .
\end{equation}

In our conventions the sigma matrices $ \sigma^{\mu}_{{\alpha} \dot{\alpha}}$,   $\bar{\sigma}^{\mu \dot{\alpha} \alpha}$  are given by
\begin{equation}
\sigma^0 = 
-\begin{pmatrix}
1 & 0 \\
0 & 1 
\end{pmatrix}, \qquad 
\sigma^1 = 
\begin{pmatrix}
0 & 1 \\
1 & 0 
\end{pmatrix}, \qquad 
\sigma^2 = 
\begin{pmatrix}
0 & -i \\
i & 0 
\end{pmatrix}, \qquad 
\sigma^3 = 
\begin{pmatrix}
1 & 0 \\
0 & -1 
\end{pmatrix}, \qquad 
\end{equation}
$\bar{\sigma} = (\sigma_0, -\sigma_1,-\sigma_2,-\sigma_3)$. Indices are raised and lowered with 
\begin{equation}
\epsilon^{\alpha \beta}=\epsilon^{\dot{\alpha} \dot{\beta}}=-\epsilon_{{\alpha} { \beta}}=-\epsilon_{\dot{\alpha} \dot{ \beta}} =\begin{pmatrix}
0 & 1 \\
-1 & 0 
\end{pmatrix}\,.
\end{equation}

The structures $\langle i, j \rangle \equiv i {R_j}_{\dot{\alpha}}
 (\bar{\sigma}_2)^{\dot{\alpha}\alpha} {L_i}_\alpha $ are preserved by the residual symmetry of the conformal frame chosen along the $x_2$ direction. Due to this it is useful to define, when working in the conformal frame, left-handed spinors $R^\alpha = i   R_{\dot{\alpha}} (\bar{\sigma}_2)^{\dot{\alpha}\alpha} = \epsilon^{\alpha \dot{\alpha}}R_{\dot{\alpha}}$. These can now be straightforwardly contracted with the $L_i$ spinors to form invariants, as is used in section \ref{TriangleSec} to define the hexagon partition function (\ref{HexagonPartition}). 

The $\langle i, j \rangle $ structures are related to the canonical covariant $V_i$, $H_{ij}$ structures used in the literature \cite{Costa:2011mg} through
 \beq\langle i, j \rangle \langle j,i \rangle  \equiv H_{ij}, \qquad \langle i, i \rangle \equiv V_i .\eeq

\section{Frames, Vertex, and Markers}
\label{explicit}

The integrability description of $\mathcal{N}=4$ SYM requires a choice of a frame, i.e. a representation for the single-trace operators in terms of magnons on top of a spin chain. Two choices are common: the string and spin-chain frame \cite{hexagons}. For most of this paper we use the spin-chain frame, in which a single trace operator made out of $L$ fundamental fields is assigned to a spin chain of length $L$. Clearly the number of fields composing an operator is not a non-perturbative notion. Moreover, it is not preserved by the dynamics: the scattering between excitations on top of the vacuum can change the chain length, deleting or introducing additional sites $\mathcal{Z}$ in the chain. These are represented by the $\mathcal{Z}^\mp$ markers. The spin-chain frame is nevertheless useful to compare with previous weak coupling results available in the literature.

The S-matrix for excitations in the spin chain frame is given by
\begin{align}
&S|\phi ^a(x) \phi^b(y)\rangle = \mathcal{A}(x,y)|\phi^{\{a}(y)\phi^{b\}}(x)\rangle + \mathcal{B}(x,y)|\phi^{ [a}(y) \phi^{b ]}(x)\rangle + \frac{1}{2}\mathcal{C}(x,y) \Sigma^{ab}\epsilon_{cd}|\mathcal{Z}^-\psi^{ c}(y) \psi^{d }(x)\rangle \nonumber \\
&S|\psi ^a(x) \psi^b(y)\rangle = \mathcal{D}(x,y)|\psi^{\{a}(y)\psi^{b\}}(x)\rangle + \mathcal{E}(x,y)|\psi^{ [a}(y) \psi^{b ]}(x)\rangle + \frac{1}{2}\mathcal{F}(x,y) \epsilon^{ab}\Sigma_{cd}|\mathcal{Z}^+\phi^{ c}(y) \phi^{d }(x)\rangle \nonumber \\
&S|\phi ^a(x) \psi^b(y)\rangle = \mathcal{G}(x,y)|\psi^{b}(y)\phi^{a}(x)\rangle + \mathcal{H}(x,y)|\phi^{a}(y)\psi^{b}(x)\rangle \nonumber\\
&S|\psi ^a(x) \phi^b(y)\rangle = \mathcal{K}(x,y)|\psi^{a}(y)\phi^{b}(x)\rangle + \mathcal{L}(x,y)|\phi^{b}(y)\psi^{a}(x)\rangle \label{smatrixkets}.
\end{align}
The S-matrix elements are expressed in terms of the Zhukovsky variables as \cite{Beisert1, Beisert2}
\begin{align}
 \mathcal{A}(x,y)&= \frac{x^{+} ( y )-x^{-} ( x )  }{x^{-} ( y )  - x^{+} ( x )}\label{A-Elem} ,\\
\nonumber \mathcal{B}(x,y)&= -1 + \frac{\left(x^{+} ( x )  - x^{+} ( y ) \right) \left(x^{-} ( x )  \left(x^{-} ( y )  - 2 x^{+} ( y ) \right) + 
      x^{+} ( x )  x^{+} ( y ) \right)}{\left(-1 + x^{-} ( x )  x^{-} ( y ) \right) \left(x^{-} ( y )  - x^{+} ( x ) \right) x^{+} ( x )  x^{+} ( y )} ,\\
\nonumber \mathcal{C}(x,y) &= \frac{2 \gamma ( x )  \gamma ( 
     y )  \left(x^{+} ( x )  - x^{+} ( y ) \right)}{\left(1 - x^{-} ( x )  x^{-} ( y ) \right) \left(x^{-} ( y )  - x^{+} ( x ) \right)}, \\
\nonumber \mathcal{D}(x,y) &= -1 ,\\
\nonumber \mathcal{E}(x,y) &=\frac{x^{+} ( y )-x^{-} ( x )}{
   x^{-} ( y )  - x^{+} ( 
     x )} + \frac{\left(x^{+} ( x )  - x^{+} ( y ) \right) \left(x^{-} ( x )  x^{-} ( y )  + 
      x^{+} ( x )  \left(x^{+} ( y )-2 x^{-} ( y ) \right)\right)}{\left(1 - x^{-} ( x )  x^{-} ( y ) \right) \left(x^{-} ( y )  - 
      x^{+} ( x ) \right) x^{+} ( x )  x^{+} ( y )}, \\
\nonumber \mathcal{F}(x,y) &= \frac{2 x^{-} ( x )  x^{-} ( 
    y )  \left(x^{-} ( x )  - x^{+} ( x ) \right) \left(x^{-} ( y )  - x^{+} ( y ) \right) \left(x^{+} ( x )  - x^{+} ( y ) \right)}{
  \gamma ( x )  \gamma ( y )  \left(-1 + x^{-} ( x )  x^{-} ( y ) \right) \left(x^{-} ( y )  - x^{+} ( x ) \right) x^{+} ( x )  x^{+} ( y )} ,\\
 \nonumber  \mathcal{G}(x,y) &= \frac{x^{+} ( y )-x^{+} ( x )}{x^{-} ( y )  - x^{+} ( x )} ,\\
\nonumber   \mathcal{H}(x,y) &= \frac{\gamma ( x )  \left(x^{+} ( y )-x^{-} ( y ) \right)}{\gamma ( y )  \left(x^{-} ( y )  - x^{+} ( x ) \right)} ,\\
\nonumber   \mathcal{L}(x,y) &= \frac{x^{-} ( y )-x^{-} ( x )}{x^{-} ( y )  - x^{+} ( x )} ,\\
\mathcal{K}(x,y) &= \frac{\gamma ( y )  \left(x^{+} ( x ) -x^{-} ( x )\right)}{\gamma ( x )  \left(x^{-} ( y )  - x^{+} ( x ) \right)}\label{K-Elem},
\end{align}
where
\begin{equation}
\gamma ( x ) = \sqrt{x^{-} ( x )  - x^{+} ( x )}. \nonumber
\end{equation}

In computing the hexagon form factors we cross all $\mathcal{D}^{\alpha \dot{\alpha}}$ excitations to the $\textbf{u}$ edge, figure (\ref{HCrossing}), break them into left- and right- fermions, and then scatter all left components with the $PSU(2|2)$ S-matrix described above. Finally we project the scattering result into the final state of right-fermions. When fermions scatter they may produce markers according to the $\mathcal{C}$ and $\mathcal{F}$ elements in (\ref{smatrixkets}). Our computation scheme is then to, as soon as a $\mathcal{Z}$ marker is created in this chain of scatterings, move it immediately to the extreme left side of the spin chain. This amounts to a translation by unit on the asymptotic wavefunctions of the excitations \cite{Beisert1, Beisert2} \beq
| \psi(u) \mathcal{Z}\rangle = \frac{\chi^+(u)}{\chi^-(u)} | \mathcal{Z} \psi(u) \rangle. \label{zmarkeraction}
\eeq
 Once on the left side of the chain, it can be ignored for the rest of the computation. The result is an extra factor in the matrix elements corresponding to the product of momentas $\frac{x^+(z)}{x^-(z)}$ for all fermions to the left of those being scattered. The result are the cumulative $\mathcal{Z}$ marker factors
\begin{align}
\phi_\mathcal{Z} (v_i,u_j) = \prod_{a=1}^{i-1}  & \prod_{b=j+1}^{J_1}  \frac{x^+(v_a)}{x^-(v_a)}\frac{x^-(u_b)}{x^+(u_b)}, \qquad  \hspace{-0.5cm}\nonumber
\phi_\mathcal{Z} (v_i,w_j) = \prod_{a=1}^{i-1}  \prod_{b=k+1}^{J_3}   \frac{x^+(v_a)}{x^-(v_a)} \frac{x^+(w_b)}{x^-(w_b)}, \qquad  \hspace{-0.5cm}\\
&\phi_\mathcal{Z} (u_j,w_k) = \prod_{a=j+1}^{J_1}  \prod_{b=1}^{J_2}  \prod_{c=k+1}^{J_3}  \frac{x^-(u_a)}{x^+(u_a)} \frac{x^+(v_b)}{x^-(v_b)} \frac{x^+(w_c)}{x^-(w_c)}. \qquad  \hspace{-0.5cm}\nonumber
\end{align}
in equation (\ref{vertex}). 

Once the $\mathcal{Z}$ marker is removed in this way, we proceed with the remaining scatterings. Note that the $x(v)$, $x(w)$ factors are inverted with respect to the $\chi(u)$ factors due to the crossed kinematics, equation (\ref{ZhuCrossing}).

In the string frame the length of the spin-chain is well defined non-perturbatively, corresponding to the $R$-charge of the dual single-trace operator. As a consequence, the S-matrix in the string frame does not produce $\mathcal{Z}$ markers and non-perturbative crossing transformations are simple. For a PSU$(2|2)^2$ bifundamental magnon excitation $\eta^{A \dot B}(z)$ crossing in the string frame simply amounts to \beq \eta^{A \dot{B}}(z) \xrightarrow{\circlearrowright} -\eta^{B \dot{A}}(z^\circlearrowright), \qquad
 \eta^{A \dot{B}}(z) \xrightarrow{\circlearrowleft} -\eta^{B \dot{A}}(z^\circlearrowleft). \label{crossingstring} \eeq Crossing in the spin-chain frame is in general complicated, requiring one to map the spin-chain frame magnons to the string frame, use  (\ref{crossingstring}), and convert back. For our purposes the procedure is trivial, since in the SL$(2,\mathbb{R})$ sector we just need to cross covariant derivatives and $\mathcal{D}_{\text{spin-chain}}^{\alpha \dot{\alpha}} = \mathcal{D}_{\text{string}}^{\alpha \dot{\alpha}}$.

\section{$\sigma$ crossing}
\label{sigmacrossing}

The crossing transformations are implemented through analytic continuation on the rapidities around the $u = \pm \frac{i}{2} + 2 g $ branch points. The S-matrix elements (\ref{A-Elem}-\ref{K-Elem}) have simple transformation rules under these monodromies, as follows from (\ref{Zhukovsky}). Clockwise $z^{\circlearrowright}$ and anti-clockwise $z^{\circlearrowleft}$ crossings are equivalent for these factors.

The dynamical factor $h(z,z')$, on the other hand, transforms non-trivially \cite{hexagons}. The necessary formulae can be derived as a consequence of the crossing equation for the BES factor,
\beq
\sigma(z^\circlearrowright, z') \sigma(z, z') = \frac{(1 - 1/x^+y^+)(1-x^-/y^+)}{(1-x^-/y^-)(1-1/x^+y^-)}
\eeq
combined with unitarity, $\sigma(z,z') \sigma(z',z) =1$. We use $x$ and $y$ to denote the Zhukovsky variable associated to $z$ and $z'$ respectively.
Some of the most useful equations are
\begin{align}
h(z^{\circlearrowright}, z') &= h(z, {z'}^{\circlearrowleft}) = \frac{1-1/x^+ y^-}{1-1/x^-y^+}\sigma(z,z')\\
h(z^{\circlearrowright}, {z'}^{\circlearrowleft}) &= \frac{y^- - x^+}{y^- - x^-}\frac{1-1/y^+x^+}{1-1/y^-x^+}\frac{1}{\sigma(z,z')}
\end{align}
which in particular imply the simplified unitarity equations referred to in figure \ref{decouplingpic}.

\section{Equal spins operators}
\label{EqualSpins}
When two or more operators have the same spin the hexagon form factors can have off-shell singularities which display an order of limits issue when going on-shell \cite{Jiang:2015bvm}.

For two spinning operators (consider $J_1=J_2=J$ and $J_3=0$), this can be easily avoided via two operations. First we cross the $v$ rapidities twice $v \to v^{\pm 4\gamma}$, which makes the matrix part of the hexagon form factor invariant and makes the kinematic pole of the diagonal limit appear only in the dynamical part. Second we use Bethe equations in the splitting factors of the $u$ particles
\begin{equation}
\omega_{\ell_{13}}(a,\bar{a}) = (-1)^{\bar{a}}\prod_{u_j \in \bar{a}} e^{-i p(u_j)\ell_{12}}\prod_{\substack{{u_i \in a}\\{i<j}}} S^{-1}(u_j,u_i) \label{CR-Weights} 
\end{equation}
which eliminates $\ell_{13}$ for $\ell_{12}$. As worked out in \cite{Jiang:2015bvm}, after summing over partitions the diagonal poles cancel resulting in the well-defined off-shell object. 

Turning on the third operator, provided that it is not equal to the other two, does not introduce new poles, so again we have well-defined off-shell object. However, for three spinning operators of equal spins ($J_1=J_2=J_3=J$) we cannot choose crossing and splitting factors in such a way that all the sets of particles satisfy the two criteria above. In other words, there is no choice of crossing and weights for the excitations $w$ that make the glued hexagon free of off-shell poles. However when going on-shell these poles must cancel with zeros coming from the matrix part and yield the physical three-point function. 

 
We were not able to come with a correct prescription for the glued hexagons that deals with the order of limit issue, therefore we avoided equal three spinning operators such as the seemingly harmless $J_1=J_2=J_3=2$ case. We believe that understanding the correct prescription is another interesting question for the hexagon formalism.

\section{From C to H's}
\label{inversion}
In this section we explain how to express the structure constant in terms of hexagon components. Our starting point is (\ref{CtoHs}), which we rewrite here as
\begin{align}
 \sum_{\ell_1,\ell_2,\ell_3} C^{J_1,J_2,J_3}_{\ell_1,\ell_2,\ell_3} \left(\frac{\langle 1,1\rangle^{J_1-\ell_2-\ell_3}\langle 2,2\rangle^{J_2-\ell_1-\ell_3}\langle 3,3\rangle^{J_3-\ell_1-\ell_2}}{\langle 2, 3 \rangle^ {-\ell_1}\langle 3,2\rangle^{-\ell_1}\langle 1, 3 \rangle^{-\ell_2} \langle 3,1\rangle^{-\ell_2}\langle 1, 2 \rangle^{-\ell_3}\langle 2,1\rangle^{-\ell_3}}\right)  =&\nonumber \\ 
 =  L_{1,\vec{\alpha}_1} L_{2,\vec{\alpha}_2}L_{3,\vec{\alpha}_3}R_{1,\vec{\beta}_1}R_{2,\vec{\beta}_2}R_{3,\vec{\beta}_3}\mathcal{H}_G^{\vec{\alpha}_1 \vec{\beta}_1;\vec{\alpha}_2 \vec{\beta}_2;\vec{\alpha}_3 \vec{\beta}_3}(\textbf{u},\textbf{v},\textbf{w})&
 \label{CtoHexApp}
\end{align}
where $\mathcal{H}_G$ are the glued hexagons.

By expanding in components and taking derivatives with respect to the spinors in both sides of the expression above we can construct a matrix $M$  that writes the polarized glued hexagons in terms of structure constants
\begin{equation}
   \mathcal{H}_G^{(\vec{\alpha}_1 \vec{\beta}_1);(\vec{\alpha}_2 \vec{\beta}_2);(\vec{\alpha}_3 \vec{\beta}_3)}  = M^{(\vec{\alpha}_1 \vec{\beta}_1);(\vec{\alpha}_2 \vec{\beta}_2);(\vec{\alpha}_3 \vec{\beta}_3)}_{\ell_1,\ell_2,\ell_3}C^{J_1,J_2,J_3}_{\ell_1,\ell_2,\ell_3}\,,
   \label{toinvert}
\end{equation}
where we emphasize that the indices of $\mathcal{H}_G$ are completely symmetrized, in accordance with
the symmetric traceless nature of the operators we are considering.

Notice that the rectangular matrix $M$ has linearly independent columns, so that $M^T M$ is invertible (since it is a grammiam matrix of linearly independent vectors). Therefore we can act on the LHS of equation (\ref{toinvert}) with the left inverse $M^+ = (M^T M)^{-1} M^T$ and find the inverted relation
\begin{equation}
 C_{J_1, J_2, J_3}^{\ell_1,\ell_2,\ell_3} = \left(M_{(\vec{\alpha}_1 \vec{\beta}_1);(\vec{\alpha}_2 \vec{\beta}_2);(\vec{\alpha}_3 \vec{\beta}_3)}^{\ell_1,\ell_2,\ell_3}\right)^{+} \mathcal{H}_G^{(\vec{\alpha}_1 \vec{\beta}_1);(\vec{\alpha}_2 \vec{\beta}_2);(\vec{\alpha}_3 \vec{\beta}_3)} 
\label{totake} 
\end{equation}
which writes the structure constants as combinations of hexagons.  Note also that the $O(3)$ invariance implies several identities between the hexagon components. These identities are simply the vanishing of the null vectors of $M^+$.

\section{Abelian $C^{\bullet\bullet\bullet}$ and Pfaffians}
\label{PfaffianApendix}

In section (\ref{abeliansection}) we presented a determinant formula for structure constants of three non protected spinning operators polarized so that two are parallel and orthogonal to the third. Here we make explicit some of the formulas and detail some steps.

First we discuss the pfaffian identity. As derived in \cite{asymptoticfourpoint}, 
\beq
\sum_{\bar{a}\subset \mathbf{u}}(-1)^{|\bar{a}|} w'(u_{\bar{a}})H(u_{\bar{a}},u_{\bar{a}}) = pf(I-\mathcal{K})_{2J_1 \times 2J_1}. \nonumber
\eeq
where the matrix $\mathcal{K}$ is defined as 
\beq
\mathcal{K}=\begin{pmatrix}
K_{11}(u,u) & K_{12}(u,u) \\
K_{21}(u,u) & K_{22} (u,u)
\end{pmatrix} \label{Ksdef}
\eeq
with 
\begin{align}
 &K_{11}(a,b)_{i j} \equiv  g^a_i/k(x^+(a_i),x^-(b_j)), \qquad K_{12}(a,b)_{i j} \equiv  -g^a_i k(x^+(a_i),x^+(b_j))/(x^+(a_i) x^+(b_j))^2, \nonumber\\
& K_{21}(a,b)_{i j} \equiv  g^a_i k(x^-(a_i),x^-(b_j))/(x^-(a_i) x^-(b_j))^2,  \qquad  K_{22}(a,b)_{i j} \equiv  - g^a_i/k(x^-(a_i),x^+(b_j)),  \nonumber
\end{align}
and where
\beq
g^u_i = k(x^+(u_i),x^-(u_i))\omega'(u_i), \qquad  k(x,y) = \frac{x-y}{1-1/xy}.
\eeq
As previously emphasized, this formula holds for any factorized function of the rapidities $w'(u_{\bar{a}} ) = \prod_{i \in \bar{a}} w'(u_i)$.

The next point we would like to clarify in this appendix is the sequence of manipulations leading to the pfaffian representation of the ${\color{blue}{\text{blue}}}$ terms. The starting point is equation (\ref{blueterms}) which reads
\begin{align}
{\color{blue}{\text{blue}}} = \sum_{\substack{ b\cup \bar{b} \subset \mathbf{v}\nonumber\\ c\cup \bar{c} \subset \mathbf{w} } }(-1)^{|\bar{b}|+|\bar{c}|} e^v_{\bar{b}}e^w_{\bar{c}}H(v_{\bar{b}},v_{\bar{b}}) H(w_{\bar{c}},w_{\bar{c}})   h(v^{\circlearrowright}_{b},\hat{u})  h(\hat{u} ,w^{\circlearrowleft}_{c})h(v^{\circlearrowright}_{b},w^{\circlearrowleft}_{c}) h(w^{\circlearrowright}_{\bar{c}} ,v^{\circlearrowleft}_{\bar{b}}). \nonumber
\end{align}
We then write
\beq
h(v^{\circlearrowright}_{b},w^{\circlearrowleft}_{c}) = \frac{h(\hat{v}^{\circlearrowright},\hat{w}^{\circlearrowleft}) h(v^{\circlearrowright}_{\bar{b}},w^{\circlearrowleft}_{\bar{c}})}{h(v^{\circlearrowright}_{\bar{b}},w^{\circlearrowleft}_{\bar{c}}) h(v^{\circlearrowright}_{\bar{b}},w^{\circlearrowleft}_{c}) h(v^{\circlearrowright}_{b},w^{\circlearrowleft}_{\bar{c}})h(v^{\circlearrowright}_{\bar{b}},w^{\circlearrowleft}_{\bar{c}})},\hspace{0.3cm} h(v^{\circlearrowright}_{b},\hat{u}) = \frac{h(\hat{v}^{\circlearrowright},\hat{u})}{h(v^{\circlearrowright}_{\bar{b}},\hat{u})}, \hspace{0.3cm} \nonumber   h(\hat{u} ,w^{\circlearrowleft}_{c}) = \frac{h(\hat{u} ,\hat{w}^{\circlearrowleft} )}{h(\hat{u} ,w^{\circlearrowleft}_{\bar{c}})}
\eeq
to end up with
\begin{align}
{\color{blue}{\text{blue}}} = & h(\hat{v}^{\circlearrowright},\hat{w}^{\circlearrowleft})  h(\hat{v}^{\circlearrowright},\hat{u}) h(\hat{u} ,\hat{w}^{\circlearrowleft} ) \sum_{\substack{ b\cup \bar{b} \subset \mathbf{v}\nonumber\\ c\cup \bar{c} \subset \mathbf{w} } }(-1)^{|\bar{b}|+|\bar{c}|} \underbrace{\left(\frac{e^v_{\bar{b}}}{h(v^{\circlearrowright}_{\bar{b}},\hat{w}^{\circlearrowleft}) h(v^{\circlearrowright}_{\bar{b}},\hat{u}) } \right)}_{\equiv\omega'(v^{\circlearrowright}_{\bar{b}})}\underbrace{\left(\frac{e^w_{\bar{c}}}{h(\hat{u} ,w^{\circlearrowleft}_{\bar{c}}) h(\hat{v}^{\circlearrowright},w^{\circlearrowleft}_{\bar{c}})}\right)}_{\equiv \omega'(w_{\bar{c}})} 
\\
& \times H(v_{\bar{b}},v_{\bar{b}}) H(w_{\bar{c}},w_{\bar{c}}) h(v^{\circlearrowright}_{\bar{b}},w^{\circlearrowleft}_{\bar{c}}) h(w^{\circlearrowright}_{\bar{c}} ,v^{\circlearrowleft}_{\bar{b}}). \nonumber
\end{align}
Here, as before, we interpret the effect of $O_1$ on operators $O_2$ and $O_3$ as a background that corrects their propagation but does not affect their interactions.
Next, we use the identity $ h(v^{\circlearrowright}_{\bar{b}},w^{\circlearrowleft}_{\bar{c}}) h(w^{\circlearrowright}_{\bar{c}} ,v^{\circlearrowleft}_{\bar{b}}) =  h(v^{\circlearrowright}_{\bar{b}},w_{\bar{c}}) h(w_{\bar{c}} ,v^{\circlearrowleft}_{\bar{b}})$ to obtain the result from the main text (\ref{blueresult}),
\beq  {\color{blue}{\text{blue}}} = h(\hat{v}^{\circlearrowright},\hat{w}^{\circlearrowleft})  h(\hat{v}^{\circlearrowright},\hat{u}) h(\hat{u} ,\hat{w}^{\circlearrowleft} )\sum_{\substack{ b\cup \bar{b} \subset \mathbf{v}\nonumber\\ c\cup \bar{c} \subset \mathbf{w} } }(-1)^{|\bar{b}|+|\bar{c}|} \omega'(v_{\bar{b}}^{\circlearrowright})\omega'(w_{\bar{c}}) H(v_{\bar{b}}^{\circlearrowright},v_{\bar{b}}^{\circlearrowright}) H(w_{\bar{c}},w_{\bar{c}})   H(v^{\circlearrowright}_{\bar{b}},w_{\bar{c}})   \nonumber. \eeq

The next step is to recognize the sum over partitions for the effective operator $O_2\cup O_3$ as a pfaffian,
\beq
\sum_{d\cup \bar{d} \subset \mathbf{z} }(-1)^{|\bar{d}|} \omega'(z_{\bar{d}}) H(z_{\bar{d}},z_{\bar{d}})  = pf(I-\mathcal{K'})_{2(J_2 + J_3) \times 2(J_2 +J_3)}. \nonumber
\eeq
Comparing with the definition $(\ref{Ksdef})$, $\mathcal{K'}$ is defined in terms of the functions $K_{x,y}(a,b)$ as
 \begin{align}\mathcal{K}'=-\begin{pmatrix}
K_{11}(v^{\circlearrowright},v^{\circlearrowright}) & K_{12}(v^{\circlearrowright},v^{\circlearrowright})   & K_{11}(v^{\circlearrowright},w)  & K_{12}(v^{\circlearrowright},w)  \\
K_{21}(v^{\circlearrowright},v^{\circlearrowright})  & K_{22}(v^{\circlearrowright},v^{\circlearrowright}) & K_{21}(v^{\circlearrowright},w)  & K_{22}(v^{\circlearrowright},w) \\
K_{11}(w,v^{\circlearrowright}) &  K_{12}(w,v^{\circlearrowright}) & K_{11}(w,w)  &  K_{12}(w,w)  \\
 K_{21}(w,v^{\circlearrowright})  & K_{22}(w,v^{\circlearrowright})  & K_{21}(w,w)  & K_{22}(w,w)  \\
\end{pmatrix} \nonumber.
\end{align}
where $g^{v^{\circlearrowright}}_i = k(x^+(v^{\circlearrowright}_i),x^-(v^{\circlearrowright}_i)) \omega'(v_i^\circlearrowright)$ and $g^{w}_i = k(x^+(w_i),x^-(w_i)) \omega'(w_i)$.

A last comment regards the choice of splitting factors in (\ref{firststep}). There we chose to cross all three sets of magnons from the front to the back hexagon through the right boundary of the cut chain. Alternative pfaffian representations can be obtained in an analogous manner for all other possible choices\footnote{e.g. left-left-left, right-right-left, etc.} of splitting factor provided one replaces (\ref{firststep}) by the appropriate expression. The only difference is the final result is the replacement of the original splitting factors $e^a_{\bar{x}}$ by their left alternatives in the pfaffian formula. As usual, the expressions differ off-shell but reproduce the same structure constants.

\end{document}